\documentclass[]{IEEEtran}

\usepackage{amssymb}
\usepackage{amsmath}
\usepackage{graphicx} 
\usepackage{epstopdf}
\usepackage{amssymb}
\usepackage{amsfonts}
\usepackage{amsthm}
\usepackage{cite}
\usepackage{bm,comment}
\usepackage{algorithm}
\usepackage{algpseudocode}

\usepackage{subeqnarray}
\usepackage{subfigure}
\usepackage{multicol}
\usepackage{multirow}
\usepackage{diagbox}
\usepackage{slashbox}
\usepackage{stfloats}
\usepackage{float}
\usepackage{color} 
\usepackage{cases}
\usepackage{multirow}
\usepackage{threeparttable}

\usepackage{lipsum}

\newtheorem{lemma}{\bf{Lemma}}

\usepackage{geometry}
\usepackage[bookmarks=true, colorlinks=true, breaklinks=true, urlcolor =black]{hyperref}

\begin{document}
\title{Robust Transceiver Design for Covert Integrated Sensing and Communications With Imperfect CSI}
\author{Yuchen Zhang, Wanli Ni, Jianquan Wang, Wanbin Tang, Min Jia,\\
Yonina C. Eldar, \emph{Fellow, IEEE}, and Dusit Niyato, \emph{Fellow, IEEE}
\vspace{-2mm}
\thanks{
Yuchen Zhang, Jianquan Wang, and Wanbin Tang are with the National Key Laboratory of Wireless Communications, University of Electronic Science and Technology of China, Chengdu 611731, China (e-mail: yc\_zhang@std.uestc.edu.cn, jqwang@uestc.edu.cn, wbtang@uestc.edu.cn).

Wanli Ni is with the State Key Laboratory of Networking and Switching Technology, Beijing University of Posts and Telecommunications, Beijing 100876, China (e-mail: charleswall@bupt.edu.cn). 

Min Jia is with the School of Electronics and Information Engineering, Harbin Institute of Technology, Harbin 150008, China (email: jiamin@hit.edu.cn).

Yonina C. Eldar is with the Faculty of Mathematics and Computer Science, Weizmann Institute of Science, Rehovot 7610001, Israel (e-mail: yonina.eldar@weizmann.ac.il).

Dusit Niyato is with the School of Computer Science and Engineering, Nanyang Technological University, Singapore 639798 (e-mail: dniyato@ntu.edu.sg).

}}

\maketitle

\vspace{-2cm}
\begin{abstract}
We propose a robust transceiver design for a covert integrated sensing and communications (ISAC) system with imperfect channel state information (CSI). Considering both bounded and probabilistic CSI error models, we formulate worst-case and outage-constrained robust optimization problems of joint transceiver beamforming and radar waveform design to balance the radar performance of multiple targets while ensuring the communications performance and covertness of the system.
The optimization problems are challenging due to the non-convexity arising from the semi-infinite constraints (SICs) and the coupled transceiver variables. 
In an effort to tackle the former difficulty, S-procedure and Bernstein-type inequality are introduced for converting the SICs into finite convex linear matrix inequalities (LMIs) and second-order cone constraints. 
A robust alternating optimization framework referred to alternating double-checking is developed for decoupling the transceiver design problem into feasibility-checking transmitter- and receiver-side subproblems, transforming the rank-one constraints into a set of LMIs, and verifying the feasibility of beamforming by invoking the matrix-lifting scheme. Numerical results are provided to demonstrate the effectiveness and robustness of the proposed algorithm in improving the performance of covert ISAC systems.
\end{abstract}


\begin{IEEEkeywords}
Integrated sensing and communications, covert communications, robust transceiver design.
\end{IEEEkeywords}

\section{Introduction}

\IEEEPARstart{D}UE to its endogenous dual-functional property, integrated sensing and communications (ISAC), envisioned as a key enabler for next-generation wireless networks, is attracting intensive research efforts from both industry and academia\cite{liu2022jsac}. Compared with traditional radar-communications coexistence, which addresses the spectrum sharing between the two separate systems, ISAC aims to unify the radar and communications functions in one platform. The integration of previously independently designed radar and communications systems can not only improve the efficiency of utilizing increasingly scarce spectrum resource, but also reduce hardware cost through platform reuse. These benefits are continuously stimulating the practical deployment of ISAC techniques such as Wi-Fi sensing and cellular networks-based environmental monitoring.

The challenge of designing an ISAC system is how to make full use of the degrees of freedom (DoFs) at both ends of the transceiver, so as to achieve a better trade-off between radar and communications performance.
As a straightforward paradigm, ISAC can be simply realized by orthogonally allocating the temporal, spectral, and spatial resources between existed radar and communications waveforms. However, due to the loose coupling of the two systems, the design DoFs cannot be fully exploited. This gives rise to the study of closely-coupled ISAC wherein a dual-functional waveform achieves the radar and communications functions simultaneously. Thus, in this paper, we focus on the transceiver design of a closely-coupled ISAC system.

\subsection{Related Works}

One of the key issues in ISAC is waveform design, which involves designing a waveform that meets the requirements of both radar and communications. This design philosophy can be categorized into three approaches: radar-centric design, communications-centric design, and joint design. The radar-centric design considers radar as the primary function and incorporates communications information into well-designed radar waveforms\cite{amin2016tsp,tianyao2020majorcom}. On the other hand, communications-centric design exploits existing communications waveforms such as orthogonal frequency division multiplexing waveform to provide radar services\cite{procIEEE2011,yixuan2022twc}. However, the random fluctuations in communications waveforms can adversely affect radar performance reliability. To strike a flexible balance between radar and communications performance, joint design has emerged, leveraging optimization techniques for waveform design.
A widely-adopted design criterion is to mimic a promising multiple input multiple output (MIMO) radar beampattern subject to the communications performance requirement by jointly optimizing the MIMO radar and MIMO communications waveforms \cite{fan2018twc,fan2018tsp,xiang2020tsp,rang2021jstsp,xiang2022jsac}. 
As a step further, the authors in \cite{fan2022tsp} leveraged Cram\'{e}r-Rao bound to measure the radar performance, thereby enabling more accurate characterization of the radar-communications performance trade-off in ISAC systems.
\begin{table*}[t]\normalsize
\caption{Contributions in contrast to the state-of-the-art}
\vspace{-2mm}
\centering
\label{contri}
\vspace{-4mm}
\begin{threeparttable}
\begin{center}
\renewcommand{\arraystretch}{1.1}
\scalebox{0.9}{
    \begin{tabular}{|c|c|c|c|c|c|c|c|}
    \hline
         & \cite{fan2018twc,fan2018tsp,xiang2020tsp,rang2021jstsp,xiang2022jsac} & \cite{bjorn2021jstsp,rang2022jsac,chen2022jsac,yiqing2022tvt,stoica2022tsp} & \cite{na2022cl} & \cite{yuanhan2022wcl}  & \cite{yonina2023arxiv,cai2023tsp} & \cite{covertISAC2023twc} & \textbf{Proposed} \\ \hline
        Multiple radar targets & $\checkmark$ & {} & {} & $\checkmark$  & $\checkmark$ & {} & $\checkmark$ \\ \hline
        Imperfect radar channel & {} & {} & {} & {} & {} & {} & $\checkmark$ \\ \hline
        Multiple communications users & $\checkmark$ & $\checkmark$ & $\checkmark$ & {}  & $\checkmark$ & {} & $\checkmark$ \\ \hline
        Imperfect communications channel & {} & {} & $\checkmark$ & {} & {} & {} & $\checkmark$ \\ \hline
        Transceiver design & {} & $\checkmark$ & $\checkmark$ & $\checkmark$  & $\checkmark$ & {} & $\checkmark$ \\ \hline
        Overt communications & $\checkmark$ & $\checkmark$ & $\checkmark$ & $\checkmark$  & $\checkmark$ & {} & $\checkmark$ \\ \hline
        Covert communications  & {} & {} & {} & {} & {} & $\checkmark$  & $\checkmark$ \\ \hline
        Multiple wardens & {} & {} & {} & {} & {} & {} & $\checkmark$ \\ \hline
    \end{tabular}
}
\end{center}
\end{threeparttable} 
\vspace{-6mm}
\end{table*}

The majority of studies have primarily focused on transmitter-side waveform design, with less emphasis on receiver-side filtering/beamforming. Recent research efforts have emerged to explore the untapped potential of radar receivers in enhancing system performance\cite{yuanhan2022wcl,chen2022jsac,yonina2023arxiv,na2022cl,yiqing2022tvt,bjorn2021jstsp,rang2022jsac,stoica2022tsp,cai2023tsp}. 
Different from heuristic metric such as beampattern error adopted in transmitter-side waveform design\cite{fan2018twc,fan2018tsp,xiang2020tsp,rang2021jstsp,xiang2022jsac}, signal-to-interference-plus-noise ratio (SINR) was usually served as a direct radar performance indicator in the literature of ISAC transceiver design. The authors in \cite{yuanhan2022wcl,chen2022jsac,yonina2023arxiv} proposed transceiver beamforming designs for multiple-target single-user, single-target multiple-user, and multple-target multiple-user ISAC systems, respectively. Specifically, to enhance the radar SINR, a Capon filter was employed in \cite{yuanhan2022wcl} and \cite{chen2022jsac} for receiver-side beamforming while steering vector of the receive antenna array was matched in \cite{yonina2023arxiv}.
Following works extended the transceiver beamforming designs to various scenarios, e.g., cooperative ISAC network\cite{na2022cl} and two-cell interfering ISAC network\cite{yiqing2022tvt}. The aforementioned designs on ISAC transceiver were based on random waveforms such that the designed statistic such as beamforming vector only reflects the average performance. However, in some circumstances, deterministic waveform design which accounts for the whole space-time block is necessary given the instantaneous waveform requirements such as constant modulus and low peak-to-average power ratio. With the pursuit of these KPIs, the authors in \cite{bjorn2021jstsp,rang2022jsac,stoica2022tsp,cai2023tsp} studied ISAC transceiver designs under the deterministic waveform regime. In particular, symbol-level precoding, jamming integration, and radar performance balance between multiple targets were further investigated in \cite{rang2022jsac,stoica2022tsp}, and \cite{cai2023tsp}, respectively. 

Due to the potential for the radar targets to be malicious and the broadcasting nature of wireless signals, it is imperative to prioritize security within the realm of ISAC systems. Along with the aforementioned efforts either utilizing the dual-functional benefit or improving the overall performance of ISAC systems, a growing body of research has started to shift its focus towards addressing security concerns. Physical layer security (PLS) aims to prevent the transmitted information from being \emph{decoded} by the eavesdropper. From the perspective of PLS, there have been various studies trying to secure ISAC systems through designing secure dual-functional waveforms\cite{jonathon2018taes, peng2022wcl} or exploiting the artificial noise/interference\cite{nanchi2021twc, xinyi2022cl, dongqi2022tvt, nanchi2022twc,noma2022tcom}.


\subsection{Motivations and Contributions}

While previous studies have shed some light on the benefits of receiver-side processing in enhancing the performance of ISAC systems, these works typically assume either perfect channel state information (CSI) or scenarios with only erroneous communications CSI. Consequently, their designs are not applicable to scenarios where both the radar and communications CSIs are imperfect, leading to a lack of robust performance.
Furthermore, certain scenarios, such as military reconnaissance signaling in battlefields and private e-health signal transmission in public, require not only the protection of transmitted information but also the concealment of the signal's existence in the first place. However, traditional PLS is insufficient to provide such a service. This highlights the need of covert communications whose goal is to safeguard the signal from being detected by a warden\cite{bash2013jsac,covert2023survey}.
The study of covert communications within the scope of the increasingly prevailing ISAC systems is still in its infancy. 
Although recent work in \cite{covertISAC2023twc} has investigated covert beamforming schemes for ISAC systems, its focus has been limited to transmitter-side design, overlooking crucial aspects related to receiver-side considerations. Furthermore, the study only considered a simplified scenario involving a single covert user and radar target. However, the generalization to a more practical scenario, where multiple covert users coexist with multiple overt/regular users and various radar targets, introduces significant challenges.

We consider transceiver design for a covert ISAC system, and propose robust schemes with imperfect radar and communications CSI. 
We compare this work with state-of-the-art schemes in Table \ref{contri} explicitly. The contributions of this paper are summarized as follows.

\begin{itemize}
\item The investigated covert ISAC system accommodates both overt and covert users, making it more versatile than previous systems that only consider one type of user. Additionally, this work represents an early attempt at transceiver design, which is a promising approach to exploiting both transmitter and receiver capabilities. 
Taking into account the channel uncertainty in both radar and communications channels, we consider two types of CSI error models: bounded and probabilistic. These models enable us to conduct a comprehensive study of worst-case and outage-constrained robust ISAC transceiver designs, while also catering to scenarios with different availability of the historical statistic for the CSI error.

\item We explore the problem of multiple radar targets coexisting with downlink communications users and signal-dependent clutters. For the purpose of balancing radar performance among each target while satisfying communication SINR and covertness of the system, we adopt a max-min optimization technique that maximizes the minimum radar SINR for each target through jointly designing the transceiver beamforming vectors and the covariance of the dedicated radar waveform. The presence of semi-infinite constraints (SICs) and coupled transceiver variables makes the optimization problems non-convex and difficult to solve directly. To overcome the SICs arising from the bounded and probabilistic CSI errors, we utilize the S-procedure and a Bernstein-type inequality to transform them into finite convex linear matrix inequalities (LMIs) and second-order cone constraints (SOCs). Furthermore, we develop an alternating optimization (AO) framework to decouple transceiver designs into feasibility-checking subproblems with respect to (w.r.t.) transmitter and receiver variables, respectively. These subproblems are recast into semi-definite programming (SDP) problems with additional rank-one constraints. Moreover, we lift the rank-one constraints into equivalent LMIs, which facilitates checking for the existence of rank-one solutions. Based on this, we propose a robust scheme called alternating double-checking (ADC) which is effective in optimizing the transceiver variables for both worst-case and outage-constrained transceiver designs. 

\item We present extensive numerical results that confirm the efficacy of our proposed schemes and provide insights that shed light on robust designs for practical systems. In particular, we demonstrate the existence of a tripartite trade-off between radar, overt and covert communications performances. We show that the transmit beampattern under radar-SINR-oriented transceiver design may not approximate the ideal radar transmit beampattern. This finding reveals the limitations of the widely-used beampattern-oriented transmitter-side design paradigm. 

\end{itemize}


\emph{Notations:} The main notations throughout this paper are clarified as follows. Lowercase, bold lowercase, and bold uppercase letters, e.g., $a$, $\mathbf{a}$, and $\mathbf{A}$, denote a scalar, vector, and matrix, respectively. $|a|$ means the absolute value of a scalar $a$ while $\|\mathbf{a} \|$ means the 2-norm of a vector $\mathbf{a}$. 
The superscripts $T$, $H$, and $*$ denote the transpose, Hermitian transpose, and conjugate transformation of a vector or matrix, respectively. 
${\text{Diag}}(\mathbf{A}_1,\mathbf{A}_2,\ldots,\mathbf{A}_N)$ denotes the matrix which is composed diagonally of matrice $\mathbf{A}_1,\mathbf{A}_2,\ldots,\mathbf{A}_N$. 
${\text{Tr}}(\mathbf{A})$, ${\text{Rank}}(\mathbf{A})$, and ${\text{Vec}}(\mathbf{A})$ mean the trace, rank, and vectorization of a matrix $\mathbf{A}$, respectively, while $\mathbf{A}\succeq \mathbf{0}$ means the matrix $\mathbf{A}$ is Hermitian and positive semi-definite. $\mathbf{0}_N$ and $\mathbf{I}_N$ denotes $N$-dimensional zero and identity matrix, respectively. ${\mathbb {E}}[\cdot]$ denotes the mathematical expectation.
$\mathcal{CN}\left(\boldsymbol{\mu},\mathbf{C}\right)$ denotes the circularly symmetric complex Gaussian (CSCG) distribution with mean $\boldsymbol{\mu}$ and covariance matrix $\mathbf{C}$.

\section{System Model}

\begin{figure}[!t]
\centering
\includegraphics[width=0.49\textwidth]{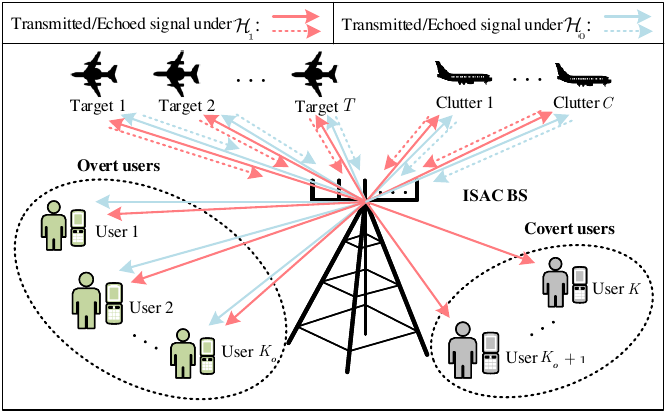}
\caption{Illustration of the covert ISAC system with $T$ radar targets, $K_{o}$ overt communications users, $K_{c}$ covert communications users, and $C$ signal-dependent clutters.}
\vspace{-2mm}
\label{ss_mod}
\end{figure}

As shown in Fig. \ref{ss_mod}, we consider an ISAC base station (BS) equipped with $M_t$ transmit antennas. The BS is constantly tracking $T$ targets which may be adversarial wardens while communicating with $K_{o}$ overt single-antenna users. In order to convey messages or commands covertly, the BS intends to communicate with other $K_{c}$ single-antenna users simultaneously while preventing the communications behaviour from being detected by the wardens. Let ${c}_{k}[n], \forall k \in \mathcal{K}_{o}\buildrel \Delta \over =\{1,\ldots,K_{o}\}$, and ${c}_{k}[n], \forall k \in \mathcal{K}_{c}\buildrel \Delta \over =\{K_{o}+1,\ldots,K_{o}+K_{c}\}$, denote the overt and covert data streams at time index $n,\forall n \in \mathcal{N}\buildrel \Delta \over =\{1,\ldots,N\}$, respectively, where $N$ denote the blocklength. Denote $\mathcal{H}_1$ and $\mathcal{H}_0$ the hypotheses that the BS transmits covert signals or not, respectively. The two-phase transmitted waveform $\mathbf{x}[n] \in \mathbb{C}^{M_t \times 1}$ is given by 
\begin{equation}\label{waveform}
\left\{
\begin{aligned}
\mathcal{H}_1:\mathbf{x}\left[n\right] &=  \mathbf{W}_{o}\mathbf{c}_{o}\left[n\right]+\mathbf{W}_{c}\mathbf{c}_{c}\left[n\right]+\mathbf{s}\left[n\right],\\
\mathcal{H}_0:\mathbf{x}\left[n\right] &= \mathbf{W}_{o}\mathbf{c}_{o}\left[n\right]+\mathbf{s}\left[n\right],
\end{aligned}
\right.
\end{equation}
where $\mathbf{W}_{o} =\left[\mathbf{w}_{1},\ldots,\mathbf{w}_{K_{o}}\right]  \in \mathbb{C}^{M_t \times K_{o}}$ and $\mathbf{W}_{c} =\left[\mathbf{w}_{K_{o}+1},\ldots,\mathbf{w}_{K_{o}+K_{c}}\right]  \in \mathbb{C}^{M_t \times K_{c}}$ consist of the transmit beamforming vectors for overt data streams $\mathbf{c}_{o}[n]=[{c}_{1}[n],\ldots,{c}_{K_{o}}[n]]^{T} \in \mathbb{C}^{K_{o} \times 1}$ and covert data streams $\mathbf{c}_{c}[n]=[{c}_{K_{o}+1}[n],\ldots,{c}_{K_{o}+K_{c}}[n]]^{T} \in \mathbb{C}^{K_{c} \times 1}$, respectively. In particular, $\mathbf{s}[n]  \in \mathbb{C}^{M_t \times 1}$ is a dedicated radar signal which can be exploited to improve the performance of the ISAC system\cite{xiang2020tsp,fan2022tsp,yonina2023arxiv}. Let $\mathcal{K}\buildrel \Delta \over =\mathcal{K}_{o} \cup \mathcal{K}_{c}$. We assume that ${c}_{k}[n],\forall k \in \mathcal{K}$, and $\mathbf{s}[n]$ are independently Gaussian distributed with ${c}_{k}[n] \sim \mathcal{CN}(0,1)$ and $\mathbf{s}[n] \sim \mathcal{CN}(\boldsymbol{0}_{M_t},\mathbf{R})$, where $\mathbf{R} \in \mathbb{C}^{M_t \times M_t}$ is the positive semidefinite covariance matrix of $\mathbf{s}[n]$.

\subsection{Communications and Radar Performances}
Let $\mathbf{h}_{{\rm{C}},k} \in \mathbb{C}^{M_t \times 1}$ denote the communications channel from the BS to the $k$-th user. Similar to \cite{xiang2020tsp,nanchi2021twc}, the received signal at the $k$-th user under $\mathcal{H}_1$ is given by
\begin{eqnarray}\label{com_sig}
\begin{aligned}
{r}_{{\rm{C}},k}\left[n\right]
=&\mathbf{h}^{H}_{{\rm{C}},k}\mathbf{w}_{k}{c}_{k}\left[n\right] + \underbrace{\sum_{i\in\mathcal{K}/k}\mathbf{h}^{H}_{{\rm{C}},k}\mathbf{w}_{i}{c}_{i}\left[n\right]}_{\text{Muti-user interference}} \\
&+  \underbrace{\mathbf{h}^{H}_{{\rm{C}},k}\mathbf{s}\left[n\right]}_{\text{Radar interference}} + {z}_{{\rm{C}},k}\left[n\right],	
\end{aligned}
\end{eqnarray}
where ${z}_{{\rm{C}},k}\left[n\right] \sim \mathcal{CN}(0,{\sigma}^2_{{\rm{C}},k})$ is the additive white Gaussian noise (AWGN). The communications performance of the $k$-th user is determined by the SINR given by\cite{nanchi2021twc}
\begin{equation}\label{com_sinr}
{\gamma}_{1,k}^{C} 
=\frac{\mathbf{h}^{H}_{{\rm{C}},k}\mathbf{w}_{k}\mathbf{w}_{k}^{H}\mathbf{h}_{{\rm{C}},k}}{\sum_{i\in\mathcal{K}/k}\mathbf{h}^{H}_{{\rm{C}},k}\mathbf{w}_{i}\mathbf{w}_{i}^{H}\mathbf{h}_{{\rm{C}},k}+\mathbf{h}^{H}_{{\rm{C}},k}\mathbf{R}\mathbf{h}_{{\rm{C}},k}+{\sigma}^2_{{\rm{C}},k}}.
\end{equation}
Similarly, under $\mathcal{H}_0$,  the communications SINR at the $k$-th user is given by
\begin{equation}\label{com_sinr H0}
{\gamma}_{0,k}^{C} =\frac{\mathbf{h}^{H}_{{\rm{C}},k}\mathbf{w}_{k}\mathbf{w}_{k}^{H}\mathbf{h}_{{\rm{C}},k}}{\sum_{i\in\mathcal{K}_{o}/k}\mathbf{h}^{H}_{{\rm{C}},k}\mathbf{w}_{i}\mathbf{w}_{i}^{H}\mathbf{h}_{{\rm{C}},k}+\mathbf{h}^{H}_{{\rm{C}},k}\mathbf{R}\mathbf{h}_{{\rm{C}},k}+{\sigma}^2_{{\rm{C}},k}}.
\end{equation}

In MIMO radar systems, the BS emits the signal and receives the echoes from the targets and clutters. Let $\theta_i,\forall i \in \mathcal{T} \buildrel \Delta \over = \{1,\ldots,T\}$, denote the angles of the targets and $\theta_i,\forall i \in \mathcal{C} \buildrel \Delta \over =\{T+1,\ldots,T+C\}$, denote the angles of the $C$ clutters. Both the communications and radar signals can be exploited as probing signals since they are perfectly known by the BS. Similar to \cite{chen2022jsac,yuanhan2022wcl}, the signal received by the colocated array with $M_r$ receive antennas is expressed as
\begin{eqnarray}\label{radar_receive_sig}
\begin{aligned}
\mathbf{r}_{{\rm{R}}}\left[n\right] = &\underbrace{\sum_{i\in\mathcal{T}}\alpha_i \mathbf{a}_{r}^{*}\left(\theta_i\right)\mathbf{a}_{t}^{H}\left(\theta_i\right) \mathbf{x}\left[n\right]}_{\text{Target echo}} \\
&+ \underbrace{\sum_{i\in\mathcal{C}}\alpha_i \mathbf{a}_{r}^{*}\left(\theta_i\right)\mathbf{a}_{t}^{H}\left(\theta_i\right) \mathbf{x}\left[n\right]}_{\text{Clutter echo}} + \mathbf{z}_{{\rm{R}}}\left[n\right],
\end{aligned}
\end{eqnarray}
where $\alpha_i$ represents the complex reflection coefficient that contains both the round-trip path loss and the radar cross-section, and $\mathbf{z}_{{\rm{R}}}\left[n\right]\sim \mathcal{CN}(\boldsymbol{0},\sigma_{{\rm{R}}}^{2}\mathbf{I}_{M_r})$ is the AWGN at the radar receiver. In addition, $\mathbf{a}_{t}\left(\theta_i\right) = [1,\ldots,e^{2\pi  (M_t-1) d \cos \theta_i /\lambda}]^{T}\in \mathbb{C}^{M_t \times 1}$ and $\mathbf{a}_{r}\left(\theta_i\right) = [1,\ldots,e^{2\pi  (M_r-1) d \cos \theta_i /\lambda}]^{T}\in \mathbb{C}^{M_r \times 1}$ are the steering vectors of the transmit and receive antenna arrays, respectively, where $d$ is the antenna interval and $\lambda$ is the wavelength. 

At the BS, the received signal is filtered by a set of receive beamforming vectors to extract the information of the targets such as location and velocity. Two groups of $T$ receive beamforming vectors should be designed to match the two-phase waveforms in \eqref{waveform}. Each vector in one of the two groups is dedicated to enhance the radar SINR w.r.t. a specific target while treating the signal echoed from other targets and clutters as interference. 
Similar to \cite{yonina2023arxiv,yiqing2022tvt,bjorn2021jstsp,na2022cl,rang2022jsac,chen2022jsac,yuanhan2022wcl}, radar SINR is adopted as the radar performance metric in the design of the ISAC transceiver. In contrast to the indirect metric used in the transmitter-side waveform design such as the beampattern error\cite{fan2018twc,fan2018tsp,xiang2020tsp,rang2021jstsp,xiang2022jsac}, SINR more directly determines radar detection and estimation performance of the target\cite{chen2022jsac}. 
Under $\mathcal{H}_1$, the $i$-th unit-power receive beamforming vector w.r.t. the target at $\theta_i$ is denoted by $\mathbf{f}_{1,i}\in \mathbb{C}^{M_r \times 1}$. Then, the corresponding radar SINR is given by\cite{yuanhan2022wcl}
\begin{align}\label{rad_sinr_h1}
&{\gamma}_{1,i}^{R}\\
=&\frac{\mathbf{f}_{1,i}^{H} \mathbf{H}_{{\rm{R}},i} \left(\sum_{k\in\mathcal{K}}\mathbf{w}_{k}\mathbf{w}_{k}^{H}+ \mathbf{R} \right) \mathbf{H}_{{\rm{R}},i}^{H}\mathbf{f}_{1,i}}{\sum_{j\in\mathcal{S}/i}\mathbf{f}_{1,i}^{H} \mathbf{H}_{{\rm{R}},j} \left(\sum_{k\in\mathcal{K}}\mathbf{w}_{k}\mathbf{w}_{k}^{H}+ \mathbf{R} \right) \mathbf{H}_{{\rm{R}},j}^{H}\mathbf{f}_{1,i} + \sigma_{{\rm{R}}}^{2}},\nonumber
\end{align}
where $\mathcal{S} \buildrel \Delta \over = \mathcal{T} \cup \mathcal{C}$ and $\mathbf{H}_{{\rm{R}},j} \buildrel \Delta \over= \alpha_i \mathbf{a}_{r}^{*}\left(\theta_i\right)\mathbf{a}_{t}^{H}\left(\theta_i\right)\in \mathbb{C}^{M_r \times M_t}$ is the radar round-trip channel. Likewise, the radar SINR under $\mathcal{H}_0$ can be expressed as
\begin{align}\label{rad_sinr_h0}
&{\gamma}_{0,i}^{R} \\
=&\frac{\mathbf{f}_{0,i}^{H} \mathbf{H}_{{\rm{R}},i}\left(\sum_{k\in\mathcal{K}_{o}}\mathbf{w}_{k}\mathbf{w}_{k}^{H}+ \mathbf{R} \right) \mathbf{H}_{{\rm{R}},i}^{H}\mathbf{f}_{0,i}}{\sum_{j\in\mathcal{S}/i}\mathbf{f}_{0,i}^{H} \mathbf{H}_{{\rm{R}},j} \left(\sum_{k\in\mathcal{K}_{o}}\mathbf{w}_{k}\mathbf{w}_{k}^{H}+ \mathbf{R} \right) \mathbf{H}_{{\rm{R}},j}^{H}\mathbf{f}_{0,i} + \sigma_{{\rm{R}}}^{2}},\nonumber
\end{align}
where $\mathbf{f}_{0,i}\in \mathbb{C}^{M_r \times 1}$ is the $i$-th unit-power receive beamforming vector w.r.t. the target at $\theta_i$. 

\subsection{Detection Performance and Covertness Constraint}

To analyze the covertness of the system, we first investigate the detection performance at the potential wardens. In addition, we consider the non-colluding scenario where the wardens cannot cooperate with each other. Thus each warden has independent detection performance. Let $\mathbf{r}_{{\rm{W}},i}=[{r}_{{\rm{W}},i}[1],{r}_{{\rm{W}},i}[2],\ldots,{r}_{{\rm{W}},i}[N]]^{T}$ be the received signal at the $i$-th warden. The hypothesis test at index $n$ is expressed as
\begin{equation}\label{hypo_test}
\left\{
\begin{aligned}
\mathcal{H}_1:{r}_{{\rm{W}},i}\left[n\right] =& \beta_i \mathbf{a}_{t}^{H}\left(\theta_i\right)\left(\mathbf{W}_{o}\mathbf{c}_{o}\left[n\right]+\mathbf{W}_{c}\mathbf{c}_{c}\left[n\right]+\mathbf{s}\left[n\right]\right) \\
&+ {z}_{{\rm{W}},i}\left[n\right],\\
\mathcal{H}_0:{r}_{{\rm{W}},i}\left[n\right] =& \beta_i \mathbf{a}_{t}^{H}\left(\theta_i\right)\left(\mathbf{W}_{o}\mathbf{c}_{o}\left[n\right]+\mathbf{s}\left[n\right]\right) + {z}_{{\rm{W}},i}\left[n\right],
\end{aligned}
\right.
\end{equation}
where $\beta_i$ and ${z}_{{\rm{W}},i}\left[n\right] \sim \mathcal{CN}(0,{\sigma}^2_{{\rm{W}},i})$ denote the corresponding path loss and AWGN, respectively. Since ${r}_{{\rm{W}},i}[n], \forall n,$ are independently identical distributed, the probability distribution functions (PDFs) of $\mathbf{r}_{{\rm{W}},i}$ under $\mathcal{H}_1$ and $\mathcal{H}_0$ can be easily derived as
\begin{align}\label{f1}
\mathbb{P}_{1,i} &\buildrel \Delta \over =\mathbb{P}\left(\mathbf{r}_{{\rm{W}},i}|\mathcal{H}_1\right)\\
&= \frac{\exp \left( -\frac{\left|\mathbf{r}_{{\rm{W}},i}\right|^2}{\mathbf{h}^H_{{\rm{W}},i}\left(\sum_{k\in\mathcal{K}}\mathbf{w}_{k}\mathbf{w}_{k}^{H}+ \mathbf{R}\right)\mathbf{h}_{{\rm{W}},i} + {\sigma}^2_{{\rm{W}},i}} \right)}{\pi^N \left(\mathbf{h}^H_{{\rm{W}},i}\left(\sum_{k\in\mathcal{K}}\mathbf{w}_{k}\mathbf{w}_{k}^{H}+ \mathbf{R}\right)\mathbf{h}_{{\rm{W}},i} + {\sigma}^2_{{\rm{W}},i}\right)^N},	\nonumber
\end{align}
and
\begin{align}\label{f0}
\mathbb{P}_{0,i} &\buildrel \Delta \over =\mathbb{P}\left(\mathbf{r}_{{\rm{W}},i}|\mathcal{H}_0\right)\\
&= \frac{\exp \left( -\frac{\left|\mathbf{r}_{{\rm{W}},i}\right|^2}{\mathbf{h}^H_{{\rm{W}},i}\left(\sum_{k\in\mathcal{K}_{o}}\mathbf{w}_{k}\mathbf{w}_{k}^{H}+ \mathbf{R}\right)\mathbf{h}_{{\rm{W}},i} + {\sigma}^2_{{\rm{W}},i}} \right)}{\pi^N \left(\mathbf{h}^H_{{\rm{W}},i}\left(\sum_{k\in\mathcal{K}_{o}}\mathbf{w}_{k}\mathbf{w}_{k}^{H}+ \mathbf{R}\right)\mathbf{h}_{{\rm{W}},i} + {\sigma}^2_{{\rm{W}},i}\right)^N},\nonumber	
\end{align}
respectively, where $\mathbf{h}_{{\rm{W}},i} \buildrel \Delta \over= \beta_i \mathbf{a}_{t}\left(\theta_i\right)\in \mathbb{C}^{M_t \times 1}$ denotes the detection channel.

Let $\mathcal{D}_1$ and $\mathcal{D}_0$ represent the decisions in support of $\mathcal{H}_1$ and $\mathcal{H}_0$, respectively. The false alarm probability is defined as $\mathbb{P}_{FA}\buildrel \Delta \over =\mathbb{P}(\mathcal{D}_1|\mathcal{H}_0)$ while the missed detection probability is defined as $\mathbb{P}_{MD}\buildrel \Delta \over =\mathbb{P}(\mathcal{D}_0|\mathcal{H}_1)$. 
Under optimal detection, the warden minimizes the detection error probability $\xi = \mathbb{P}_{FA}+\mathbb{P}_{MD}$ with the minimum denoted by $\xi^*$. The covertness constraint of the system is expressed as $\xi^* \buildrel \Delta \over =\mathbb{P}_{FA}+\mathbb{P}_{MD} \ge 1-\epsilon$, where $\epsilon \in [0,1]$ is the covertness constant. 
The expression of $\xi^*$ involves complicated gamma functions which leads to a mathematically intractable optimization problem. To circumvent this difficulty, we surrogate $\xi^*$ by its lower bound via leveraging the Kullback-Leibler divergence defined as $\mathcal{D}(\mathbb{P}_{1,i}\left|\right|\mathbb{P}_{0,i})\buildrel \Delta \over =\int_{\mathbf{r}_{{\rm{W}},i}|\mathcal{H}_1}{\mathbb{P}_{1,i}\ln (\frac{\mathbb{P}_{1,i}}{\mathbb{P}_{0,i}})\mathrm{d}\mathbf{r}_{{\rm{W}},i}}$. Based on Pinsker's inequality\cite{zack2023twc}, we have $\xi^*  \ge 1-\sqrt{\frac{\mathcal{D}(\mathbb{P}_{1,i}\left|\right|\mathbb{P}_{0,i})}{2}}$. It is straightforward to verify that the original covertness constraint is fulfilled provided $\mathcal{D}(\mathbb{P}_{1,i}||\mathbb{P}_{0,i})\leq 2\epsilon^2$. Furthermore, by substituting \eqref{f1} and \eqref{f0} into the expression of $\mathcal{D}(\mathbb{P}_{1,i}||\mathbb{P}_{0,i})$, we have $\mathcal{D}\left(\mathbb{P}_{1,i}||\mathbb{P}_{0,i}\right)= N f \left(\frac{\mathbf{h}^H_{{\rm{W}},i}\left(\sum_{k\in\mathcal{K}_{c}}\mathbf{w}_{k}\mathbf{w}_{k}^{H}\right)\mathbf{h}_{{\rm{W}},i} }{\mathbf{h}^H_{{\rm{W}},i}\left(\sum_{k\in\mathcal{K}_{o}}\mathbf{w}_{k}\mathbf{w}_{k}^{H}+ \mathbf{R}\right)\mathbf{h}_{{\rm{W}},i} + {\sigma}^2_{{\rm{W}},i}}\right)$,
where the function $f \left(x\right)=x-\ln\left(1+x\right)$ is monotonically increasing for $x \ge 0$. Then, the covertness constraint $\mathcal{D}(\mathbb{P}_{1,i}||\mathbb{P}_{0,i})\leq 2\epsilon^2$ can be simplified as 
\begin{eqnarray}\label{cc_u}
\begin{aligned}
\mathbf{h}^H_{{\rm{W}},i}\mathbf{\Xi}\mathbf{h}_{{\rm{W}},i}  \le \eta{\sigma}^2_{{\rm{W}},i},
\end{aligned}
\end{eqnarray} 
where $\mathbf{\Xi}\buildrel \Delta \over=\sum_{k\in\mathcal{K}_{c}}\mathbf{w}_{k}\mathbf{w}_{k}^{H} - \eta\left(\sum_{k\in\mathcal{K}_{o}}\mathbf{w}_{k}\mathbf{w}_{k}^{H}+ \mathbf{R}\right)\in \mathbb{C}^{M_t \times M_t}$ and $\eta \ge 0$ is the solution of $f \left(x\right) = \frac{2\epsilon^2}{N}$. Specifically, when $\epsilon = 0$, $\eta=0$.

\subsection{Channel Error Model}

The considered covert ISAC system involves three types of channels including the communications channels $\mathbf{h}_{{\rm{C}},k}, \forall k \in \mathcal{K}$, detection channels $\mathbf{h}_{{\rm{W}},i}, \forall i \in \mathcal{T} $, and radar round-trip channel $\mathbf{H}_{{\rm{R}},j}, \forall j \in \mathcal{S}$. 
In practice, due to inevitable errors arising from estimation or quantization, the CSI of the communications channel cannot be perfectly known at the BS\cite{chanError2014tsp,gui2020tsp}. 
Additionally, the directions of the targets are unlikely to be perfectly known, introducing certain degree of ambiguities in the angles\cite{nanchi2021twc,nanchi2022twc}. Moreover, the RCSs of the targets are also unlikely to be perfectly estimated. Both factors lead to imperfect radar CSI. 
Instead of assuming perfect CSI or solely considering imperfect communications CSI as studies summarized in Table \ref{contri}, we aim to investigate the scenario where both the radar and communications channels are imperfectly known at the BS.
The CSIs of the three kinds of channels are expressed as the sum of the estimated value and estimation error, i.e., $\mathbf{h}_{{\rm{C}},k} = \widetilde{\mathbf{h}}_{{\rm{C}},k} + \Delta \mathbf{h}_{{\rm{C}},k}$, $\mathbf{h}_{{\rm{W}},i} = \widetilde{\mathbf{h}}_{{\rm{W}},i} + \Delta \mathbf{h}_{{\rm{W}},i}$, and $\mathbf{H}_{{\rm{R}},j} = \widetilde{\mathbf{H}}_{{\rm{R}},j} + \Delta \mathbf{H}_{{\rm{R}},j}$, respectively. The accuracy of the CSI has a significant impact on the performance of the system. In the following, we introduce two types of the CSI error model.

\begin{itemize}
\item \emph{Bounded CSI error}: The norms of the CSI errors are bounded by known constants $e_{{\rm{C}},k}^{2}$, $e_{{\rm{W}},i}^{2}$, and $e_{{\rm{R}},j}^{2}$, respectively, i.e.,
\begin{subequations}\label{be}
\begin{align}
&\Delta \mathbf{h}_{{\rm{C}},k} \in {\rm{C}}_k, \forall k \in \mathcal{K}, \label{be-a}\\
&\Delta \mathbf{h}_{{\rm{W}},i} \in {\rm{W}}_i, \forall i \in \mathcal{T}, \label{be-b}\\
&\text{Vec}\left(\Delta \mathbf{H}_{{\rm{R}},j}\right) \in {\rm{R}}_j, \forall j \in \mathcal{S},\label{be-c}
\end{align}	
\end{subequations}
where ${\rm{C}}_k \buildrel \Delta \over =\{\Delta \mathbf{h}_{{\rm{C}},k}|\left\|\Delta \mathbf{h}_{{\rm{C}},k}\right\|^{2} \le e_{{\rm{C}},k}^{2} \}$, ${\rm{W}}_i \buildrel \Delta \over =\{\Delta \mathbf{h}_{{\rm{W}},i}|\left\|\Delta \mathbf{h}_{{\rm{W}},i}\right\|^{2} \le e_{{\rm{W}},i}^{2} \}$, and ${\rm{R}}_j \buildrel \Delta \over =\{\text{Vec}(\Delta \mathbf{H}_{{\rm{R}},j})|\left\|\text{Vec}(\Delta \mathbf{H}_{{\rm{R}},j})\right\|^{2} \le e_{{\rm{R}},j}^{2} \}$.
\item \emph{Probabilistic CSI error}: The CSI error vectors are characterized by a CSCG distribution with a known covariance matrix, i.e., 
\begin{subequations}\label{pe}
\begin{align}
&\Delta \mathbf{h}_{{\rm{C}},k} \sim \mathcal{CN}\left(\boldsymbol{0},\mathbf{E}_{{\rm{C}},k}\right),\forall k \in \mathcal{K}, \label{pe-a}\\
&\Delta \mathbf{h}_{{\rm{W}},i} \sim \mathcal{CN}\left(\boldsymbol{0},\mathbf{E}_{{\rm{W}},i}\right),\forall i \in \mathcal{T},  \label{pe-b}\\
&\text{Vec}\left(\Delta \mathbf{H}_{{\rm{R}},j}\right) \sim \mathcal{CN}\left(\boldsymbol{0},\mathbf{E}_{{\rm{R}},j}\right), \forall j \in \mathcal{S}, \label{pe-c}
\end{align}	
\end{subequations}
where $\mathbf{E}_{{\rm{C}},k}\in \mathbb{C}^{M_t \times M_t}$, $\mathbf{E}_{{\rm{W}},i}\in \mathbb{C}^{M_t \times M_t}$, and $\mathbf{E}_{{\rm{R}},j}\in \mathbb{C}^{M_t M_r \times M_t M_r}$ are positive semidefinite error covariance matrices.
\end{itemize}


\section{Worst-Case Robust Design Under Bounded Error Model}
In this section, we investigate the worst-case robust transceiver design under bounded error model. To be specific, in the two-phase transmission, we aim to maximize the minimum radar SINR among all targets subject to fulfilling the communications rate requirement and covertness constraint over all possible channel realizations. The formulated problem is non-convex, which is recast into a SDP problem with rank-one constraints. Then, we present a matrix-lifting approach to reformulate the rank-one constraints, and propose an AO-based algorithm to solve this problem.


\subsection{Problem Formulation}

For the two-phase transmission, we aim to maximize the minimum radar SINR among all targets subject to fulfilling the communications SINR requirement over all possible channel realizations. We can verify that ${\gamma}_{1,k}^{C}\le {\gamma}_{0,k}^{C}$ from \eqref{com_sinr} and \eqref{com_sinr H0}. The communications SINR and transmit power are limited by those under $\mathcal{H}_1$. Therefore, the optimization problem w.r.t. the transceiver beamforming vectors and the radar covariance matrix is formulated as
\begin{subequations}\label{p1}
\begin{align}
&\mathop {\max }\limits_{\mathbf{w}_{k}, \mathbf{R}, \mathbf{f}_{1,i}, \mathbf{f}_{0,i}} \mathop {\min}\limits_{i \in \mathcal{T}} \mathop {\min}\limits_{\text{Vec}\left(\Delta \mathbf{H}_{{\rm{R}},j}\right) \in {\rm{R}}_j}\; \left\{{\gamma}_{1,i}^{R},{\gamma}_{0,i}^{R}\right\}\label{p1-a}\\
{\rm{s.t.}}\;
&{\gamma}_{1,k}^{C} \ge \Gamma_k,\forall \Delta \mathbf{h}_{{\rm{C}},k} \in {\rm{C}}_k, k \in \mathcal{K},\label{p1-b}\\
&\mathbf{h}^H_{{\rm{W}},i}\mathbf{\Xi}\mathbf{h}_{{\rm{W}},i}  \le \eta{\sigma}^2_{{\rm{W}},i},\forall \Delta \mathbf{h}_{{\rm{W}},i} \in {\rm{W}}_i, i \in \mathcal{T},\label{p1-c}\\
&{\text{Tr}}\left(\sum_{k\in\mathcal{K}}\mathbf{w}_{k}\mathbf{w}_{k}^{H}+ \mathbf{R}\right)\le P,\label{p1-d}\\
&{\text{Tr}}\left(\mathbf{f}_{1,i}\mathbf{f}_{1,i}^{H}\right)=1,{\text{Tr}}\left(\mathbf{f}_{0,i}\mathbf{f}_{0,i}^{H}\right)=1,\forall i \in \mathcal{T},\label{p1-e}\\
&\mathbf{R}\succeq \mathbf{0},
\end{align}
\end{subequations}
where $\Gamma_1,\ldots,\Gamma_{K_{o}}$ and $\Gamma_{K_{o}+1},\ldots,\Gamma_{K_{o}+K_{c}}$ denote the targeted SINRs for the overt and covert communications, respectively, and $P$ is the power budget. Note that when both the radar and communications CSIs are perfectly known and the receive beamforming is fixed to match the steering vector of each of the targets, \eqref{p1} is similar to the radar SINR balance problem in \cite{yonina2023arxiv}, which is readily solved by designing the transmitter-side waveform in an iterative manner. However, the approach in \cite{yonina2023arxiv} cannot be applied to \eqref{p1} due to the non-convex SICs induced by CSI errors and the covertness constraint. Moreover, since the receive beamforming is not determined straightforwardly, the coupled nature of the transmit and receive variables in the expressions of radar SINR further complicates the transceiver design.


\subsection{Problem Reformulation}

To tackle the SICs \eqref{p1-b} and \eqref{p1-c}, we introduce the following lemma.

\begin{lemma}
(S-Procedure\cite{gui2020tsp}): Define the quadratic functions w.r.t. $\mathbf{x} \in \mathbb{C}^{M \times 1}$ as 
\begin{eqnarray}\label{s_lemma_quad}
\begin{aligned}
f_{m}\left(\mathbf{x}\right) = 	\mathbf{x}^{H} \mathbf{A}_{m}\mathbf{x} + 2 \text{Re}\left\{\mathbf{b}_{m}^{H} \mathbf{x}\right\} + c_{m}, m =1,2,	
\end{aligned}	
\end{eqnarray}
where $\mathbf{A}_{m} \in \mathbb{C}^{M \times M}$, $\mathbf{b}_{m} \in \mathbb{C}^{M \times 1}$, and $c_{m} \in \mathbb{R}$. The condition $f_{1} \le 0 \Rightarrow f_{2} \le 0$ holds if and only if there exists a variable $\omega \ge 0$ such that
\begin{eqnarray}\label{s_lemma_matrix}
\begin{aligned}
\omega	
\begin{bmatrix}
\mathbf{A}_{1} & \mathbf{b}_{1}\\
\mathbf{b}_{1}^{H} & c_{1}
\end{bmatrix}
-
\begin{bmatrix}
\mathbf{A}_{2} & \mathbf{b}_{2}\\
\mathbf{b}_{2}^{H} & c_{2}
\end{bmatrix}
\succeq \mathbf{0}_{M+1}.
\end{aligned}	
\end{eqnarray}	
\end{lemma}

\newcounter{MYtempeqncn}
\begin{figure*}[!b]
\hrulefill
\normalsize
\setcounter{MYtempeqncn}{\value{equation}}
\setcounter{equation}{20}
\begin{eqnarray}\label{rad_sinr_h1_kron}
\begin{aligned}
{\gamma}_{1,i}^{R} =\frac{ \text{Vec}^{H}\left( \mathbf{H}_{{\rm{R}},i}\right) \left(\left(\sum_{k\in\mathcal{K}}\mathbf{W}_{k}+ \mathbf{R} \right)^{*}\otimes \mathbf{F}_{1,i}\right)  \text{Vec}\left( \mathbf{H}_{{\rm{R}},i}\right)}{\sum_{j\in\mathcal{S}/i}\text{Vec}^{H}\left( \mathbf{H}_{{\rm{R}},j}\right) \left(\left(\sum_{k\in\mathcal{K}}\mathbf{W}_{k}+ \mathbf{R} \right)^{*}\otimes \mathbf{F}_{1,i}\right)  \text{Vec}\left( \mathbf{H}_{{\rm{R}},j}\right) + \sigma_{{\rm{R}}}^{2}},
\end{aligned}
\end{eqnarray}
\begin{eqnarray}\label{rad_sinr_h0_kron}
\begin{aligned}
{\gamma}_{0,i}^{R} =\frac{ \text{Vec}^{H}\left( \mathbf{H}_{{\rm{R}},i}\right) \left(\left(\sum_{k\in\mathcal{K}_{o}}\mathbf{W}_{k}+ \mathbf{R}\right)^{*}\otimes \mathbf{F}_{0,i}\right)  \text{Vec}\left( \mathbf{H}_{{\rm{R}},i}\right)}{\sum_{j\in\mathcal{S}/i}\text{Vec}^{H}\left( \mathbf{H}_{{\rm{R}},j}\right) \left(\left(\sum_{k\in\mathcal{K}_{o}}\mathbf{W}_{k}+ \mathbf{R}\right)^{*}\otimes \mathbf{F}_{0,i}\right)  \text{Vec}\left( \mathbf{H}_{{\rm{R}},j}\right) + \sigma_{{\rm{R}}}^{2}},
\end{aligned}
\end{eqnarray}
\setcounter{equation}{\value{MYtempeqncn}}
\end{figure*} 
  
Note that $\Delta \mathbf{h}_{{\rm{C}},k} \in {\rm{C}}_k$ leads to $\Delta \mathbf{h}_{{\rm{C}},k}^{H}\Delta \mathbf{h}_{{\rm{C}},k} \le e_{{\rm{C}},k}^{2}$. By substituting $\mathbf{h}_{{\rm{C}},k} = \widetilde{\mathbf{h}}_{{\rm{C}},k} + \Delta \mathbf{h}_{{\rm{C}},k}$ into \eqref{p1-b}, ${\gamma}_{1,k}^{C} \ge \Gamma_k$ can be transformed into 
\begin{eqnarray}\label{def_com_1}
\begin{aligned}
\Delta \mathbf{h}_{{\rm{C}},k}^{H} \mathbf{\Psi}_{k}\Delta \mathbf{h}_{{\rm{C}},k}&+2 \mathbf{\Psi}_{k}\widetilde{\mathbf{h}}_{{\rm{C}},k}-\Gamma_k {\sigma}^2_{{\rm{C}},k}\\
&+ \widetilde{\mathbf{h}}_{{\rm{C}},k}^{H} \mathbf{\Psi}_{k} \widetilde{\mathbf{h}}_{{\rm{C}},k} \ge 0,
\end{aligned}	
\end{eqnarray}
where $ \mathbf{\Psi}_{k} \buildrel \Delta \over= \mathbf{W}_{k}-\sum_{q=1,q \neq k}^{K_{o}+K_{c}}\Gamma_k \mathbf{W}_{q}-\Gamma_k\mathbf{R} \in \mathbb{C}^{M_t \times M_t}$ with $\mathbf{W}_{k} \buildrel \Delta \over =\mathbf{w}_{k}\mathbf{w}_{k}^{H} \in \mathbb{C}^{M_t \times M_t}$. 
According to Lemma 1, \eqref{p1-b} is equivalent to 
\begin{eqnarray}\label{com_sinr_matrix}
\begin{aligned}
&\begin{bmatrix}
\mu_{k} \mathbf{I} + \mathbf{\Psi}_{k}  & \mathbf{\Psi}_{k}\widetilde{\mathbf{h}}_{{\rm{C}},k}\\
\widetilde{\mathbf{h}}_{{\rm{C}},k}^{H} \mathbf{\Psi}_{k} & \widetilde{\mathbf{h}}_{{\rm{C}},k}^{H} \mathbf{\Psi}_{k} \widetilde{\mathbf{h}}_{{\rm{C}},k} - \Gamma_k {\sigma}^2_{{\rm{C}},k} - \mu_{k}e_{{\rm{C}},k}^{2}
\end{bmatrix}
\succeq \mathbf{0},	\\ 
&\mu_{k}\ge 0,\forall k \in \mathcal{K},
\end{aligned}	
\end{eqnarray}
%
where $\mu_{k}$ is the introduced auxiliary variable.

Similarly, the constraint \eqref{p1-c} can be transformed into 
\begin{eqnarray}\label{covertness_constraint}
\begin{aligned}
&\begin{bmatrix}
\phi_{i} \mathbf{I} - \mathbf{\Xi} & -\mathbf{\Xi}\widetilde{\mathbf{h}}_{{\rm{W}},i}\\
-\widetilde{\mathbf{h}}_{{\rm{W}},i}^{H}\mathbf{\Xi} & \eta{\sigma}^2_{{\rm{W}},i}-\widetilde{\mathbf{h}}_{{\rm{W}},i}^{H}\mathbf{\Xi}\widetilde{\mathbf{h}}_{{\rm{W}},i}-\phi_{i} e_{{\rm{W}},i}^{2} 
\end{bmatrix}
\succeq \mathbf{0},\\
&\phi_{i}\ge 0,\forall i \in \mathcal{T}, \label{ccm+}
\end{aligned}
\end{eqnarray}
where $\phi_{i}$ is the introduced auxiliary variable. 

Next, in order to handle the complicated max-min objective function, we introduce an auxiliary variable $t \buildrel \Delta \over = \mathop {\min}\limits_{i \in \mathcal{T}} \mathop {\min}\limits_{\text{Vec}(\Delta \mathbf{H}_{{\rm{R}},j}) \in {\rm{R}}_j}\{{\gamma}_{1,i}^{R},{\gamma}_{0,i}^{R}\}$ as the lower bound of radar SINR. Define $\mathbf{F}_{1,i} \buildrel \Delta \over =\mathbf{f}_{1,i} \mathbf{f}_{1,i}^{H} \in \mathbb{C}^{M_r \times M_r}$ and  $\mathbf{F}_{0,i} \buildrel \Delta \over =\mathbf{f}_{0,i} \mathbf{f}_{0,i}^{H} \in \mathbb{C}^{M_r \times M_r}$. The objective function \eqref{p1-a} can be reformulated as
\begin{subequations}\label{p2}
\begin{align}
&\qquad\mathop {\max }\limits_{\mathbf{W}_{k},\mathbf{R}, \mathbf{F}_{1,i}, \mathbf{F}_{0,i},t} \; t\label{p2-a}\\
&\qquad{\rm{s.t.}}\; 
{\gamma}_{1,i}^{R}  \ge   t,{\gamma}_{0,i}^{R}  \ge  t,\nonumber\\
&\;\;\;\;\;\;\;\;\;\;\;\;\;\;\text{Vec}\left(\Delta \mathbf{H}_{{\rm{R}},j}\right)  \in  {\rm{R}}_j,\forall i  \in  \mathcal{T},j  \in  \mathcal{S}. \label{p2-b}
\end{align}
\end{subequations}

By applying linear manipulations, the radar SINRs under $\mathcal{H}_1$ and $\mathcal{H}_0$ can be transformed into \eqref{rad_sinr_h1_kron} and \eqref{rad_sinr_h0_kron} at the bottom of this page, respectively.
Moreover, we define $\widetilde{\mathbf{g}}_{\rm{R}} \buildrel \Delta \over= [\text{Vec}(\widetilde{\mathbf{H}}_{{\rm{R}},1}), \ldots, \text{Vec}(\widetilde{\mathbf{H}}_{{\rm{R}},T+C})]^{T} \in \mathbb{C}^{(T+C)M_t M_r \times 1}$, $\Delta \mathbf{g}_{\rm{R}} \buildrel \Delta \over= [\text{Vec}(\Delta \mathbf{H}_{{\rm{R}},1}), \ldots, \text{Vec}(\Delta \mathbf{H}_{{\rm{R}},T+C})]^{T} \in \mathbb{C}^{(T+C)M_t M_r \times 1}$, and ${\rm{R}} \buildrel \Delta \over =\{\Delta \mathbf{g}_{\rm{R}}|\left\|\Delta \mathbf{g}_{\rm{R}}\right\|^{2}  \le e_{{\rm{R}}}^{2} \buildrel \Delta \over= \sum_{j\in\mathcal{S}} e_{{\rm{R}},j}^{2} \}$.
Then, \eqref{p2-b} is recast as the quadratic forms 
\addtocounter{equation}{2}
\begin{eqnarray}\label{rad_sinr_matrix}
\begin{aligned}
\Delta \mathbf{g}_{\rm{R}}^{H}\mathbf{X}_{1,i}\Delta\mathbf{g}_{\rm{R}}+&2 \text{Re}\left\{\left(\mathbf{X}_{1,i}\widetilde{\mathbf{g}}_{\rm{R}}\right)^{H} \Delta \mathbf{g}_{\rm{R}}\right\}+\widetilde{\mathbf{g}}_{\rm{R}}^{H}\mathbf{X}_{1,i}\widetilde{\mathbf{g}}_{\rm{R}}\\
+&\sigma_{{\rm{R}}}^{2} \le 0,\\
\Delta \mathbf{g}_{\rm{R}}^{H}\mathbf{X}_{0,i}\Delta\mathbf{g}_{\rm{R}}+&2 \text{Re}\left\{\left(\mathbf{X}_{0,i}\widetilde{\mathbf{g}}_{\rm{R}}\right)^{H} \Delta \mathbf{g}_{\rm{R}}\right\}+\widetilde{\mathbf{g}}_{\rm{R}}^{H}\mathbf{X}_{0,i}\widetilde{\mathbf{g}}_{\rm{R}}\\
+&\sigma_{{\rm{R}}}^{2} \le 0,\Delta \mathbf{g}_{\rm{R}} \in {\rm{R}},\forall i \in \mathcal{T},
\end{aligned}
\end{eqnarray}
where 
\begin{eqnarray}\label{def_big_X1}
\begin{aligned}
\mathbf{X}_{1,i} \buildrel \Delta \over= \text{Diag}&\left(\left(\sum_{k\in\mathcal{K}}\mathbf{W}_{k}+ \mathbf{R} \right)^{*}\otimes \mathbf{F}_{1,i},\ldots, \mathbf{D}_{1,i},\right.\\
&\left.\ldots,\left(\sum_{k\in\mathcal{K}}\mathbf{W}_{k}+ \mathbf{R}\right)^{*}\otimes \mathbf{F}_{1,i}\right)
\end{aligned}	
\end{eqnarray}
and
\begin{eqnarray}\label{def_big_X0}
\begin{aligned}
\mathbf{X}_{0,i} \buildrel \Delta \over= \text{Diag}&\left(\left(\sum_{k\in\mathcal{K}_{o}}\mathbf{W}_{k}+ \mathbf{R}\right)^{*}\otimes \mathbf{F}_{0,i},\ldots, \mathbf{D}_{0,i},\right.\\
&\left.\ldots,\left(\sum_{k\in\mathcal{K}_{o}}\mathbf{W}_{k}+ \mathbf{R}\right)^{*}\otimes  \mathbf{F}_{0,i}\right).	
\end{aligned}	
\end{eqnarray}
Additionally, the $i$-th matrices of $\mathbf{X}_{1,i}$ and $\mathbf{X}_{0,i}$ are given by $\mathbf{D}_{1,i} \buildrel \Delta \over =-(\sum_{k\in\mathcal{K}}\mathbf{W}_{k}+ \mathbf{R} )^{*}\otimes \mathbf{F}_{1,i}/t \in \mathbb{C}^{M_t M_r \times M_t M_r}$ and $\mathbf{D}_{0,i} \buildrel \Delta \over = -(\sum_{k\in\mathcal{K}_{o}}\mathbf{W}_{k}+ \mathbf{R})^{*}\otimes \mathbf{F}_{0,i}/t \in \mathbb{C}^{M_t M_r \times M_t M_r}$, respectively. 
Again, based on the above formulations, we can leverage Lemma 1 to transform \eqref{rad_sinr_matrix} into LMIs as
\begin{align}\label{rad_sinr_final_matrix}
&\begin{bmatrix}
\varpi_{1,i} \mathbf{I} - \mathbf{X}_{1,i} & -\mathbf{X}_{1,i}\widetilde{\mathbf{g}}_{\rm{R}}\\
-\widetilde{\mathbf{g}}_{\rm{R}}^{H}\mathbf{X}_{1,i} & -\widetilde{\mathbf{g}}_{\rm{R}}^{H}\mathbf{X}_{1,i}\widetilde{\mathbf{g}}_{\rm{R}}-\sigma_{{\rm{R}}}^{2}-\varpi_{1,i} \sum_{j\in\mathcal{S}} e_{{\rm{R}},j}^{2}
\end{bmatrix}
\succeq \mathbf{0},\nonumber\\
&\varpi_{1,i}\ge 0,	\\	
&\begin{bmatrix}
\varpi_{0,i} \mathbf{I} - \mathbf{X}_{0,i} & -\mathbf{X}_{0,i}\widetilde{\mathbf{g}}_{\rm{R}}\\
-\widetilde{\mathbf{g}}_{\rm{R}}^{H}\mathbf{X}_{0,i} & -\widetilde{\mathbf{g}}_{\rm{R}}^{H}\mathbf{X}_{0,i}\widetilde{\mathbf{g}}_{\rm{R}}-\sigma_{{\rm{R}}}^{2}-\varpi_{0,i} \sum_{j\in\mathcal{S}} e_{{\rm{R}},j}^{2}
\end{bmatrix}
\succeq \mathbf{0},\nonumber\\
&\varpi_{0,i} \ge 0, \forall i \in \mathcal{T},\nonumber 
\end{align}
where $\varpi_{1,i}$ and $\varpi_{0,i}$ are the introduced auxiliary variables.
Now, we can rewrite \eqref{p1} as
\begin{subequations}\label{p3}
\begin{align}
&\mathop {\max }\limits_{\mathbf{W}_{k},\mathbf{R}, \mathbf{F}_{1,i}, \mathbf{F}_{0,i},\mu_{k},\phi_{i},\varpi_{1,i},\varpi_{0,i},t} \; t\label{p3-a}\\
{\rm{s.t.}}\; 
&{\text{Tr}}\left(\sum_{k\in\mathcal{K}}\mathbf{W}_{k}+ \mathbf{R}\right)\le P,\label{p3-b}\\
&{\text{Tr}}\left(\mathbf{F}_{1,i}\right)=1,{\text{Tr}}\left(\mathbf{F}_{0,i}\right)=1,\forall i \in \mathcal{T},\label{p3-c}\\
&\mathbf{W}_{k}\succeq \mathbf{0},\forall k \in \mathcal{K}, \mathbf{R}\succeq \mathbf{0},\label{p3-d}\\
&\mathbf{F}_{1,i}\succeq \mathbf{0}, \mathbf{F}_{0,i}\succeq \mathbf{0},\forall i \in \mathcal{T},\label{p3-e}\\
&\text{Rank}\left(\mathbf{W}_{k}\right)=1, \forall k \in \mathcal{K}, \label{p3-f}\\
&\text{Rank}\left(\mathbf{F}_{1,i}\right)=1,\text{Rank}\left(\mathbf{F}_{0,i}\right)=1,\forall i \in \mathcal{T}, \label{p3-g}\\
&\eqref{com_sinr_matrix},\eqref{covertness_constraint},\eqref{rad_sinr_final_matrix}.
\end{align}
\end{subequations}

Omitting the non-convex rank-one constraints \eqref{p3-f} and \eqref{p3-g}, the above problem is relaxed as
\begin{eqnarray}\label{p4}
\begin{aligned} 
&\mathop {\max }\limits_{\mathbf{W}_{k}, \mathbf{R}, \mathbf{F}_{1,i}, \mathbf{F}_{0,i},\mu_{k},\phi_{i},\varpi_{1,i},\varpi_{0,i},t} \;\;\; t \\
{\rm{s.t.}}\; 
&\eqref{com_sinr_matrix},\eqref{covertness_constraint},\eqref{rad_sinr_final_matrix},\eqref{p3-b},\eqref{p3-c},\eqref{p3-d},\eqref{p3-e}.
\end{aligned}	
\end{eqnarray}
Problem \eqref{p4} is still non-convex due to coupled transmit and receive variables, as well as the objective value $t$ in $\mathbf{D}_{1,i}$ and $\mathbf{D}_{0,i}$. In the following, we propose an AO-based approach to solve this problem efficiently.

\subsection{Proposed Transceiver Design Framework}
Given the transmit (receive) variables and $t$, the subproblem w.r.t. the receive (transmit) variables is a convex feasibility-checking SDP problem. Based on this property, we employ the AO framework to solve \eqref{p4}. Specifically, we optimize the transmit and receive variables in an alternate manner. Define the lower and upper bound of $t$ as $\underline{t}$ and $\overline{t}$, respectively. In the $l$-th iteration, we first deal with the subproblem w.r.t. the transmit variables while fixing the receive variables to the values of the last iteration, i.e., $\mathbf{F}_{1,i}^{\left(l-1\right)}$ and $\mathbf{F}_{0,i}^{\left(l-1\right)}, \forall i \in \mathcal{T}$. The convex feasibility-checking problem is formulated as
\begin{subequations}\label{p4-Tx}
\begin{align} 
{\text{find} } \; & {\mathbf{W}_{k}, \mathbf{R},\mu_{k},\phi_{i},\varpi_{1,i},\varpi_{0,i}} \\
{\rm{s.t.}}\; 
&t = {t}_{\text{tx}}^{\left(l\right)},\mathbf{F}_{1,i}=\mathbf{F}_{1,i}^{\left(l-1\right)}, \mathbf{F}_{0,i}=\mathbf{F}_{0,i}^{\left(l-1\right)}, \forall i \in \mathcal{T},\label{p4-Tx-b}\\
&\eqref{com_sinr_matrix},\eqref{covertness_constraint},\eqref{rad_sinr_final_matrix},\eqref{p3-b},\eqref{p3-d}, 
\end{align}	
\end{subequations}
where ${t}_{\text{tx}}^{\left(l\right)}$ is the given objective value. The initial lower and upper bounds for ${t}_{\text{tx}}^{\left(l\right)}$ are given by $\underline{t}_{\text{tx}}^{\left(l\right)} = {t}^{\left(l-1\right)}$ and $\overline{t}_{\text{tx}}^{\left(l\right)} = \overline{t}$, respectively, where ${t}^{\left(l-1\right)}$ is the objective value of the last iteration. Then, we try to improve ${t}_{\text{tx}}^{\left(l\right)}$ by checking the feasibility of \eqref{p4-Tx} iteratively. To be exact, in each inner loop, we update the objective value by ${t}_{\text{tx}}^{\left(l\right)}=\frac{\underline{t}_{\text{tx}}^{\left(l\right)}+\overline{t}_{\text{tx}}^{\left(l\right)}}{2}$. If feasible, we update the lower bound by $\underline{t}_{\text{tx}}^{\left(l\right)} = {t}_{\text{tx}}^{\left(l\right)}$, and store $\mathbf{W}_{k}^{\left(l\right)}, \forall k \in \mathcal{K}$, and $\mathbf{R}^{\left(l\right)}$. Otherwise, we update the upper bound by $\underline{t}_{\text{tx}}^{\left(l\right)} = {t}_{\text{tx}}^{\left(l\right)}$. We repeat this process until $\overline{t}_{\text{tx}}^{\left(l\right)} - \underline{t}_{\text{tx}}^{\left(l\right)} \le \delta$, where $\delta$ is a predetermined tolerance. Then, we renew ${t}^{\left(l\right)} = {t}_{\text{tx}}^{\left(l\right)}$.

Similarly, for the subproblem w.r.t. the receive variables, we initialize the lower and upper bounds of the objective value as $\underline{t}_{\text{rx}}^{\left(l\right)} = {t}^{\left(l\right)}$ and $\overline{t}_{\text{rx}}^{\left(l\right)} = \overline{t}$. The corresponding convex feasibility-checking problem is given by
\begin{subequations}\label{p4-Rx}
\begin{align} 
{\text{find} } \;  &{\mathbf{F}_{1,i},\mathbf{F}_{0,i}, \varpi_{1,i},\varpi_{0,i}} \\
{\rm{s.t.}}\;
&t = {t}_{\text{rx}}^{\left(l\right)},\mathbf{W}_{k} = \mathbf{W}_{k}^{\left(l\right)},\forall k \in \mathcal{K},\label{p4-Rx-b}\\
&\eqref{rad_sinr_final_matrix},\eqref{p3-c},\eqref{p3-e}, 
\end{align}	
\end{subequations}
where ${t}_{\text{rx}}^{\left(l\right)}$ is the given objective value. Again, ${t}_{\text{rx}}^{\left(l\right)}$, $\underline{t}_{\text{rx}}^{\left(l\right)}$, and $\overline{t}_{\text{rx}}^{\left(l\right)}$ are iteratively updated by the bisection approach until the convergence condition $\overline{t}_{\text{rx}}^{\left(l\right)} - \underline{t}_{\text{rx}}^{\left(l\right)} \le \delta$ is met. Then, we renew ${t}^{\left(l\right)} = {t}_{\text{rx}}^{\left(l\right)}$, $\underline{t}_{\text{tx}}^{\left(l\right)} = {t}^{\left(l\right)}$, and $\overline{t}_{\text{tx}}^{\left(l\right)} = \overline{t}$, and enter the $(l+1)$-th iteration until the maximum iteration number $L$ is reached. Note that the convergence of the algorithm could be theoretically guaranteed given sufficiently large $L$, as we analyze later. In fact, it usually takes no more than 4 outer iterations before the algorithm saturates, as we show in the numerical results.

To achieve convexity, the rank-one constraints \eqref{p3-f} and \eqref{p3-g} are omitted. Hence the output $\mathbf{W}_{k}$ and $\mathbf{F}_{i}$ from above steps may not be rank-one. 
To circumvent this problem, we integrate the rank-one constraints into the subproblems and propose a double-checked AO framework, i.e., ADC to make sure the rank-one property of corresponding transmit and receive variables in the following. To begin with, we present the following lemma \cite{huang2023tsp} which can be used to lift the rank-one constraint into a more tractable matrix form. 
\begin{lemma}
For a positive semi-definite Hermitian matrix $\mathbf{A} \in \mathbb{C}^{M \times M}$, the condition $\text{Rank}\left(\mathbf{A}\right)=1$ is equivalent to the following conditions
\begin{subequations}\label{lemma2}
\begin{align}
&\text{Tr}\left(\mathbf{A}\mathbf{B}\right)-2v-	\text{Tr}\left(\mathbf{V}\right) \ge 0,\text{Tr}\left(\mathbf{B}\right)=1,\\
&\mathbf{V} - \mathbf{A} + v \mathbf{I}_{M}\succeq \mathbf{0}_{M}, \mathbf{B} \succeq \mathbf{0}_{M},\mathbf{V} \succeq \mathbf{0}_{M},\label{lemma2-b}
\end{align}	
\end{subequations}
where $v$ and the Hermitian matrices $\mathbf{B},\mathbf{V} \in \mathbb{C}^{M \times M}$ are the introduced auxiliary variables.
\end{lemma}

Incorporating constraint \eqref{p3-f} via Lemma 2, \eqref{p4-Tx} is recast as 
\begin{subequations}\label{p4-Tx-rank}
\begin{align} 
{\text{find} } \;  &{\mathbf{W}_{k},\mathbf{R},\mathbf{B}_{k},\mathbf{V}_{k},\mu_{k},\phi_{i},\varpi_{1,i},\varpi_{0,i}}, v_{k}\\
{\rm{s.t.}}\;
&\text{Tr}\left(\mathbf{W}_{k}\mathbf{B}_{k}\right)\!-\!2 v_{k}\!-	\!\text{Tr}\left(\mathbf{V}_{k}\right)\! \ge \!0,\text{Tr}\left(\mathbf{B}_{k}\right)\!=\!1,\label{p4-Tx-rank-b}\\
&\mathbf{V}_{k}\! -\! \mathbf{W}_{k}\! + \! v_{k} \mathbf{I}\!\succeq \!\mathbf{0},\mathbf{B}_{k}\! \succeq \!\mathbf{0},\mathbf{V}_{k}\! \succeq \!\mathbf{0},\forall k \!\in \!\mathcal{K}, \label{p4-Tx-rank-c}\\
&\eqref{com_sinr_matrix},\eqref{covertness_constraint},\eqref{rad_sinr_final_matrix},\eqref{p3-b},\eqref{p3-d},\eqref{p4-Tx-b}, 
\end{align}	
\end{subequations}
where $ v_{k}$ and $\mathbf{B}_{k}$, $\mathbf{V}_{k} \in \mathbb{C}^{M_t  \times M_t }$ are the introduced auxiliary variables. During the inner loop w.r.t. the transmit variables, whenever \eqref{p4-Tx} is feasible with ${t}_{\text{tx}}^{\left(l\right)}$, it is necessary to double-check if feasible transmit variables can also be found by \eqref{p4-Tx-rank} with the same ${t}_{\text{tx}}^{\left(l\right)}$. If not, we should not use ${t}_{\text{tx}}^{\left(l\right)}$ to update $\underline{t}_{\text{tx}}^{\left(l\right)}$ but $\overline{t}_{\text{tx}}^{\left(l\right)}$, i.e., to reduce the upper bound. Because transmit variables with rank-one property are not found even if \eqref{p4-Tx} is feasible. 
Note that the only difficulty for double-checking is that \eqref{p4-Tx-rank} is non-convex due to the coupled $\mathbf{W}_{k}$ and $\mathbf{B}_{k}$ in constraint \eqref{p4-Tx-rank-b}. Nevertheless, we can overcome this difficulty in an iterative manner. Specifically, by fixing $\mathbf{W}_{k}$ in constraint \eqref{p4-Tx-rank-b} to the value of the last loop, \eqref{p4-Tx-rank} becomes convex and can be solved to update $\mathbf{W}_{k}$ iteratively. If the value of $\mathbf{W}_{k}$ converges, i.e., the mean-square error of the values of two consecutive iterations is less than a given threshold $\tau$ in $J$ iterations, the double-checking procedure is successful, and the corresponding rank-one transmit variables are found. Otherwise, the double-checking procedure fails, we should use ${t}_{\text{tx}}^{\left(l\right)}$ to update $\overline{t}_{\text{tx}}^{\left(l\right)}$ instead of $\underline{t}_{\text{tx}}^{\left(l\right)}$. The initial value of $\mathbf{W}_{k}$ is determined by the feasible solutions of \eqref{p4-Tx} with ${t}_{\text{tx}}^{\left(l\right)}$.


\renewcommand{\algorithmicrequire}{\textbf{Input:}}
\renewcommand{\algorithmicensure}{\textbf{Output:}}
\begin{algorithm}[!t]
\caption{Double-Checking for the Worst-Case Transmitter (Receiver) Design }
\label{algo 1}       %
\begin{algorithmic}[1]
\State \textbf{Input}: $\mathbf{F}_{1,i}^{\left(l-1\right)}$, $\mathbf{F}_{0,i}^{\left(l-1\right)}$, $\underline{t}_{\text{tx}}^{\left(l\right)}$, $\overline{t}_{\text{tx}}^{\left(l\right)}$ ($\mathbf{W}_{k}^{\left(l-1\right)}$, $\underline{t}_{\text{rx}}^{\left(l\right)}$, $\overline{t}_{\text{rx}}^{\left(l\right)}$), $\delta$, $\tau$, and $J$;
\While {$\overline{t}_{\text{tx}}^{\left(l\right)} - \underline{t}_{\text{tx}}^{\left(l\right)}>\delta$ ($\overline{t}_{\text{rx}}^{\left(l\right)} - \underline{t}_{\text{rx}}^{\left(l\right)}>\delta$)}
\State Set ${t}_{\text{tx}}^{\left(l\right)} = (\underline{t}_{\text{tx}}^{\left(l\right)}+\overline{t}_{\text{tx}}^{\left(l\right)})/2$ (${t}_{\text{rx}}^{\left(l\right)} = (\underline{t}_{\text{rx}}^{\left(l\right)}+\overline{t}_{\text{rx}}^{\left(l\right)})/2$);
\State Let $\mathbf{F}_{1,i}=\mathbf{F}_{1,i}^{\left(l-1\right)}$, $\mathbf{F}_{0,i}=\mathbf{F}_{0,i}^{\left(l-1\right)}$, $t = {t}_{\text{tx}}^{\left(l\right)}$ ($\mathbf{W}_{k}=\mathbf{W}_{k}^{\left(l-1\right)}$, $\mathbf{R}=\mathbf{R}^{\left(l-1\right)}$, $t = {t}_{\text{rx}}^{\left(l\right)}$), and check the feasibility of \eqref{p4-Tx} (check \eqref{p4-Rx}).
\If {feasible}
\State {Set $\mathbf{W}_{k}^{\left(l,0\right)} = \mathbf{W}_{k}^{\left(l\right)}$ ($\mathbf{F}_{1,i}^{\left(l,0\right)} = \mathbf{F}_{1,i}^{\left(l\right)}$ and $\mathbf{F}_{0,i}^{\left(l,0\right)} = \mathbf{F}_{0,i}^{\left(l\right)}$);}
\For {$j = 1:J$}
\State {Check the feasibility of \eqref{p4-Tx-rank} via setting $\mathbf{W}_{k}$ in \eqref{p4-Tx-rank-b} as $\mathbf{W}_{k}^{\left(l,j-1\right)}$ (\eqref{p4-Rx-rank} via setting $\mathbf{F}_{1,i}$ in \eqref{p4-Rx-rank-b} and $\mathbf{F}_{0,i}$ in \eqref{p4-Rx-rank-d} as $\mathbf{F}_{1,i}^{\left(l,j-1\right)}$ and $\mathbf{F}_{0,i}^{\left(l,j-1\right)}$, respectively);}
\If {feasible}
\State {Store $\mathbf{W}_{k}^{\left(l,j\right)}$ ($\mathbf{F}_{1,i}^{\left(l,j\right)}$ and $\mathbf{F}_{0,i}^{\left(l,j\right)}$);}
\If {$\|\mathbf{W}_{k}^{\left(l,j\right)}-\mathbf{W}_{k}^{\left(l,j-1\right)}\|>\tau$ ($\|\mathbf{F}_{1,i}^{\left(l,j\right)}-\mathbf{F}_{1,i}^{\left(l,j-1\right)}\|>\tau$ or $\|\mathbf{F}_{0,i}^{\left(l,j\right)}-\mathbf{F}_{0,i}^{\left(l,j-1\right)}\|>\tau$)}
\State{Go to line 19;}
\Else
\State{Set $\underline{t}_{\text{tx}}^{\left(l\right)}={t}_{\text{tx}}^{\left(l\right)}$ ($\underline{t}_{\text{rx}}^{\left(l\right)}={t}_{\text{rx}}^{\left(l\right)}$) and break;}
\EndIf
\Else
\State{Set $\overline{t}_{\text{tx}}^{\left(l\right)}={t}_{\text{tx}}^{\left(l\right)}$ ($\overline{t}_{\text{rx}}^{\left(l\right)}={t}_{\text{rx}}^{\left(l\right)}$);}
\EndIf 
\State{$j = j +1$;}
\EndFor
\State{Set $\overline{t}_{\text{tx}}^{\left(l\right)}={t}_{\text{tx}}^{\left(l\right)}$ ($\overline{t}_{\text{rx}}^{\left(l\right)}={t}_{\text{rx}}^{\left(l\right)}$);}
\Else
\State {Set $\overline{t}_{\text{tx}}^{\left(l\right)}={t}_{\text{tx}}^{\left(l\right)}$ ($\overline{t}_{\text{rx}}^{\left(l\right)}={t}_{\text{rx}}^{\left(l\right)}$);}
\EndIf 
\EndWhile
\State Output $\mathbf{W}_{k} = \mathbf{W}_{k}^{\left(l\right)}$, $\mathbf{R} = \mathbf{R}^{\left(l\right)}$, ${t}_{\text{tx}}^{\left(l\right)}$ ($\mathbf{F}_{1,i} = \mathbf{F}_{1,i}^{\left(l\right)}$, $\mathbf{F}_{0,i} = \mathbf{F}_{0,i}^{\left(l\right)}$, ${t}_{\text{rx}}^{\left(l\right)}$).
\end{algorithmic}
\end{algorithm}

Likewise, by incorporating constraint \eqref{p3-g}, receiver-side problem \eqref{p4-Rx} is transformed into 
\begin{subequations}\label{p4-Rx-rank}
\begin{align} 
{\text{find} }\;  &{\mathbf{F}_{1,i}, \mathbf{F}_{0,i}, \mathbf{U}_{1,i}, \mathbf{U}_{0,i}, \mathbf{Z}_{1,i}, \mathbf{Z}_{0,i}, \varpi_{1,i}, \varpi_{0,i}, u_{1,i}, u_{0,i}} \\
{\rm{s.t.}}\;
&\text{Tr} \left(\mathbf{F}_{1,i}\mathbf{U}_{1,i}\right) - 2u_{1,i} -	 \text{Tr} \left(\mathbf{Z}_{1,i}\right)  \ge  0, \text{Tr} \left(\mathbf{U}_{1,i}\right) = 1 ,\label{p4-Rx-rank-b}\\
&\mathbf{Z}_{1,i}  -  \mathbf{U}_{1,i}  +  u_{1,i} \mathbf{I} \succeq  \mathbf{0}_{M_r} , \mathbf{U}_{1,i}  \succeq  \mathbf{0} , \mathbf{Z}_{1,i}  \succeq  \mathbf{0} , \label{p4-Rx-rank-c}\\
&\text{Tr} \left(\mathbf{F}_{0,i}\mathbf{U}_{0,i}\right) - 2u_{0,i} - 	\text{Tr} \left(\mathbf{Z}_{0,i}\right)  \ge  0 , \text{Tr} \left(\mathbf{U}_{0,i}\right) = 1 , \label{p4-Rx-rank-d}\\
&\mathbf{Z}_{0,i}  -  \mathbf{U}_{0,i}  +  u_{0,i} \mathbf{I} \succeq  \mathbf{0} , \mathbf{U}_{0,i}  \succeq  \mathbf{0} , \mathbf{Z}_{0,i}  \succeq  \mathbf{0} , \forall i  \in  \mathcal{T} , \label{p4-Rx-rank-e}\\
&\eqref{rad_sinr_final_matrix},\eqref{p3-c},\eqref{p3-e},\eqref{p4-Rx-b},
\end{align}	
\end{subequations}
where $u_{1,i}$, $u_{0,i}$, and $\mathbf{U}_{1,i}, \mathbf{U}_{0,i},\mathbf{Z}_{1,i},\mathbf{Z}_{0,i} \in \mathbb{C}^{M_r  \times M_r }$ are the introduced auxiliary variables. The receiver-side double-checking is similar to the transmitter-side one. We do not repeat it for brevity. Algorithm \ref{algo 1} details the steps of double-checking for both transmitter and receiver in parallel. Based on this, the robust scheme ADC for the worst-case transceiver design under bounded error model is summarized in Algorithm \ref{algo 2}. The output transmit and receive variables $\mathbf{W}_{k}$, $\mathbf{F}_{1,i}$, and $\mathbf{F}_{0,i}$ are guaranteed to be rank-one. By simply using the eigenvalue decomposition (ED), we can get the corresponding optimized beamforming vectors. 

We remark that the initial receive variables can be determined by $\mathbf{F}_{1,i}^{\left(0\right)} = \mathbf{I}_{M_r}/M_r$ and $\mathbf{F}_{0,i}^{\left(0\right)} = \mathbf{I}_{M_r}/M_r$. Besides, the lower bound $\underline{t}$ should be chosen as a sufficiently small value such that \eqref{p4} with $\mathbf{F}_{1,i}=\mathbf{F}_{1,i}^{\left(0\right)}$, $\mathbf{F}_{1,i}=\mathbf{F}_{1,i}^{\left(0\right)}, \forall i \in \mathcal{T}$, and $t = \underline{t}$ is feasible. Otherwise, the optimization will fail because no feasible solutions can be found even in the first iteration of the inner loop w.r.t. the transmit variables. Meanwhile, the upper bound $\overline{t}$ should be chosen as a sufficiently large value such that \eqref{p4} is always infeasible at the first iteration of the inner loop w.r.t. both transmit and receive variables. Otherwise, the algorithm will end before reaching the stationary point. We further remark that although the feasible receive variables are not necessarily rank-one after the first check, simulations show that it is almost always the case. This indicates that most of the time we could obtain the rank-one $\mathbf{F}_{1,i}$ and $\mathbf{F}_{0,i}$ directly without undergoing the second check.

\renewcommand{\algorithmicrequire}{\textbf{Input:}}
\renewcommand{\algorithmicensure}{\textbf{Output:}}
\begin{algorithm}[!t]
\caption{ADC for the Worst-Case Transceiver Design}
\label{algo 2}       %
\begin{algorithmic}[1]
\State \textbf{Input}: Initial receive variables $\mathbf{F}_{1,i}^{\left(0\right)}$ and $\mathbf{F}_{0,i}^{\left(0\right)}$, $\underline{t}$, $\overline{t}$, $\delta$, $L$, $\tau$, and $J$;
\State Set $l=1$ and ${t}^{\left(0\right)}= \underline{t}$;
\For {$l \le L$}
\State Set $\underline{t}_{\text{tx}}^{\left(l\right)} = {t}^{\left(l-1\right)}$ and $\overline{t}_{\text{tx}}^{\left(l\right)} = \overline{t}$, optimize transmitter via the transmitter-side Algorithm \ref{algo 1};
\State Set ${t}^{\left(l\right)} = {t}_{\text{tx}}^{\left(l\right)}$;
\State Set $\underline{t}_{\text{rx}}^{\left(l\right)} = {t}^{\left(l\right)}$ and $\overline{t}_{\text{rx}}^{\left(l\right)} = \overline{t}$, optimize receiver via the receiver-side Algorithm \ref{algo 1};
\State Set ${t}^{\left(l\right)} = {t}_{\text{rx}}^{\left(l\right)}$ and $l=l+1$; 
\EndFor
\State Output $\mathbf{W}_{k} = \mathbf{W}_{k}^{\left(l\right)}$, $\mathbf{R} = \mathbf{R}^{\left(L\right)}$, $\mathbf{F}_{1,i} = \mathbf{F}_{1,i}^{\left(L\right)}$, $\mathbf{F}_{0,i} = \mathbf{F}_{0,i}^{\left(L\right)}$, $t = t^{\left(L\right)}$.
\end{algorithmic}
\end{algorithm}


\section{Outage-Constrained Robust Design Under Probabilistic Error Model}
In practice, the worst-case design under bounded error model may be too strict and fails to capture the historical statistic of the CSI error\cite{chanError2014tsp}. To address this issue, in this section, we will focus on the probabilistic CSI error model and investigate the corresponding outage-constrained robust transceiver design. Similar to the worst-case design, in the following, we first formulate the problem and transform the SICs therein into convex LMIs and SOCs. Then, we leverage the AO framework to solve this problem.


\subsection{Problem Formulation}
The outage probabilities for radar SINR, communications SINR, and covertness constraint are defined as $\rho_{{\rm{R}},i}$, $\rho_{{\rm{C}},k}$, and $\rho_{{\rm{W}},i}$, respectively. Besides, as in the worst-case design, we lift the transmit and receive beamforming vectors as matrices. The optimization problem to maximize the minimum outage-constrained radar SINR among all targets, denoted by $t$, is formulated as
\begin{subequations}\label{p5}
\begin{align}
&\mathop {\max }\limits_{\mathbf{W}_{k},  \mathbf{R}, \mathbf{F}_{1,i}, \mathbf{F}_{0,i},t} t\label{p5-a}\\
{\rm{s.t.}}\;
&\text{Pr}\left\{ \gamma_{1,i}^{R} \ge t \right\}\ge 1- \rho_{{\rm{R}},i},\text{Pr}\left\{ \gamma_{0,i}^{R} \ge t \right\}\ge 1- \rho_{{\rm{R}},i},\nonumber\\
&\Delta \mathbf{g}_{\rm{R}} \sim \mathcal{CN}\left(\boldsymbol{0},\mathbf{E}_{{\rm{R}}}\right),\forall i \in \mathcal{T},\label{p5-b}\\
&\text{Pr}\left\{ \gamma_{1,k}^{C} \ge \Gamma_k\right\}\ge 1- \rho_{{\rm{C}},k},\nonumber\\
&\Delta \mathbf{h}_{{\rm{C}},k} \sim \mathcal{CN}\left(\boldsymbol{0},\mathbf{E}_{{\rm{C}},k}\right),\forall k \in \mathcal{K},\label{p5-c}\\
&\text{Pr}\left\{ \eqref{cc_u}  \right\}\ge 1- \rho_{{\rm{W}},i},\nonumber\\
&\Delta \mathbf{h}_{{\rm{W}},i} \sim \mathcal{CN}\left(\boldsymbol{0},\mathbf{E}_{{\rm{W}},i}\right),\forall i \in \mathcal{T},\label{p5-d}\\
&\eqref{p3-b},\eqref{p3-c},\eqref{p3-d},\eqref{p3-e},\eqref{p3-f},\eqref{p3-g}, 
\end{align}
\end{subequations}
where \eqref{p5-c} and \eqref{p5-d} denote the outage-constrained constraints for communications SINR and covertness of the system, respectively, and $\mathbf{E}_{{\rm{R}}}\buildrel \Delta \over=\text{Diag}(\mathbf{E}_{{\rm{R}},1},\ldots,\mathbf{E}_{{\rm{R}},T+C})\in \mathbb{C}^{(T+C)M_t M_r \times (T+C)M_t M_r}$. Due to the probabilistic errors, constraints \eqref{p5-b}-\eqref{p5-d} have no simple close-form expressions, which makes the above problem computationally prohibitive. 

\vspace{-2mm}

\subsection{Problem Reformulation and Solving}
To deal with constraints \eqref{p5-b}-\eqref{p5-d}, we introduce the following useful lemma.
\begin{lemma}
(Bernstein-Type Inequality\cite{chanError2014tsp}): Given random variable $\mathbf{x} \sim \mathcal{CN}(\boldsymbol{0},\mathbf{I}_{M})$, $\mathbf{A} \in \mathbb{C}^{M \times M}$, $\mathbf{b} \in \mathbb{C}^{M \times 1}$, $s \in \mathbb{R}$, and $\rho \in [0,1]$. The sufficient conditions for
\begin{eqnarray}\label{bern_quad}
\begin{aligned}
\text{Pr}\left\{\mathbf{x}^{H} \mathbf{A}\mathbf{x} + 2 \text{Re}\left\{\mathbf{b} \mathbf{x}\right\} + s \ge 0 \right\} \ge 1- \rho,	
\end{aligned}	
\end{eqnarray}
are presented in the following constraints 
\begin{eqnarray}\label{bern_cond}
\begin{aligned}
&\text{Tr}\left(\mathbf{A}\right)-\sqrt{2\ln\left(\frac{1}{\rho}\right)}x +\ln\left(\rho\right)y+s \ge 0,\\
&\sqrt{\left\|\mathbf{A}\right\|_{\text{F}}^{2}+2\left\|\mathbf{b}\right\|^{2}} \le x,  y\mathbf{I}_{M} + \mathbf{A} \succeq \mathbf{0}_{M},  y \ge 0, \label{bern_cond_d}	
\end{aligned}	
\end{eqnarray}
where $x$ and $y$ are the introduced auxiliary variables.
\end{lemma}

We first handle the radar SINR constraints. Define $\Delta \widehat{\mathbf{g}}_{\rm{R}} \buildrel \Delta \over= \mathbf{E}_{{\rm{R}}}^{-\frac{1}{2}}\Delta \mathbf{g}_{\rm{R}}$. Similar to the formulation of \eqref{rad_sinr_matrix}, the events presented in the probabilistic constraint \eqref{p5-b} can be transformed into
\begin{eqnarray}\label{rad_sinr_pe_matrix}
\begin{aligned}
&\Delta \widehat{\mathbf{g}}_{\rm{R}}^{H}\mathbf{E}_{{\rm{R}}}^{\frac{1}{2}}\mathbf{X}_{1,i}\mathbf{E}_{{\rm{R}}}^{\frac{1}{2}}\Delta\widehat{\mathbf{g}}_{\rm{R}}+2 \text{Re}\left\{\left(\mathbf{E}_{{\rm{R}}}^{\frac{1}{2}}\mathbf{X}_{1,i}\widetilde{\mathbf{g}}_{\rm{R}}\right)^{H} \Delta \widehat{\mathbf{g}}_{\rm{R}}\right\}+\\
&\widetilde{\mathbf{g}}_{\rm{R}}^{H}\mathbf{X}_{1,i}\widetilde{\mathbf{g}}_{\rm{R}}+\sigma_{{\rm{R}}}^{2} \le 0,\\
&\Delta \widehat{\mathbf{g}}_{\rm{R}}^{H}\mathbf{E}_{{\rm{R}}}^{\frac{1}{2}}\mathbf{X}_{0,i}\mathbf{E}_{{\rm{R}}}^{\frac{1}{2}}\Delta\widehat{\mathbf{g}}_{\rm{R}}+2 \text{Re}\left\{\left(\mathbf{E}_{{\rm{R}}}^{\frac{1}{2}}\mathbf{X}_{0,i}\widetilde{\mathbf{g}}_{\rm{R}}\right)^{H} \Delta \widehat{\mathbf{g}}_{\rm{R}}\right\}+\\
&\widetilde{\mathbf{g}}_{\rm{R}}^{H}\mathbf{X}_{0,i}\widetilde{\mathbf{g}}_{\rm{R}}+\sigma_{{\rm{R}}}^{2} \le 0,\Delta \widehat{\mathbf{g}}_{\rm{R}} \sim \mathcal{CN}\left(\boldsymbol{0},\mathbf{I}\right),\forall i \in \mathcal{T}.
\end{aligned}
\end{eqnarray}
Subsequently, by Lemma 2, the sufficient conditions for constraint \eqref{p5-b} are given by
\begin{eqnarray}\label{rad_sinr_pe_cond}
\begin{aligned}
&\text{Tr}\left(\mathbf{E}_{{\rm{R}}}^{\frac{1}{2}}\mathbf{X}_{1,i}\mathbf{E}_{{\rm{R}}}^{\frac{1}{2}}\right)+\sqrt{2\ln\left(\frac{1}{\rho_{{\rm{R}},i}}\right)}x_{1,i}^{R} -\ln\left(\rho_{{\rm{R}},i}\right)y_{1,i}^{R}+\\
&\widetilde{\mathbf{g}}_{\rm{R}}^{H}\mathbf{X}_{1,i}\widetilde{\mathbf{g}}_{\rm{R}}+\sigma_{{\rm{R}}}^{2} \le 0, \\
&\sqrt{\left\|\mathbf{E}_{{\rm{R}}}^{\frac{1}{2}}\mathbf{X}_{1,i}\mathbf{E}_{{\rm{R}}}^{\frac{1}{2}}\right\|_{\text{F}}^{2}+2\left\|\mathbf{E}_{{\rm{R}}}^{\frac{1}{2}}\mathbf{X}_{1,i}\widetilde{\mathbf{g}}_{\rm{R}}\right\|^{2}} \le  x_{1,i}^{R}, \\
&y_{1,i}^{R}\mathbf{I} + \mathbf{E}_{{\rm{R}}}^{\frac{1}{2}}\mathbf{X}_{1,i}\mathbf{E}_{{\rm{R}}}^{\frac{1}{2}} \succeq \mathbf{0}, y_{1,i}^{R} \ge 0,	 \\
&\text{Tr}\left(\mathbf{E}_{{\rm{R}}}^{\frac{1}{2}}\mathbf{X}_{0,i}\mathbf{E}_{{\rm{R}}}^{\frac{1}{2}}\right)+\sqrt{2\ln\left(\frac{1}{\rho_{{\rm{R}},i}}\right)}x_{0,i}^{R} -\ln\left(\rho_{{\rm{R}},i}\right)y_{0,i}^{R}+\\
&\widetilde{\mathbf{g}}_{\rm{R}}^{H}\mathbf{X}_{0,i}\widetilde{\mathbf{g}}_{\rm{R}}+\sigma_{{\rm{R}}}^{2} \le 0,	\label{rad_sinr_pe_cond_a}\\
&\sqrt{\left\|\mathbf{E}_{{\rm{R}}}^{\frac{1}{2}}\mathbf{X}_{0,i}\mathbf{E}_{{\rm{R}}}^{\frac{1}{2}}\right\|_{\text{F}}^{2}+2\left\|\mathbf{E}_{{\rm{R}}}^{\frac{1}{2}}\mathbf{X}_{0,i}\widetilde{\mathbf{g}}_{\rm{R}}\right\|^{2}} \le x_{0,i}^{R}, \\
&y_{0,i}^{R}\mathbf{I} + \mathbf{E}_{{\rm{R}}}^{\frac{1}{2}}\mathbf{X}_{0,i}\mathbf{E}_{{\rm{R}}}^{\frac{1}{2}} \succeq \mathbf{0}, y_{0,i}^{R} \ge 0,\forall i \in \mathcal{T},
\end{aligned}
\end{eqnarray}
where $x_{1,i}^{R}$, $x_{0,i}^{R}$, $y_{1,i}^{R}$, and $y_{0,i}^{R}$ are the introduced auxiliary variables. 

Likewise, let $\Delta \widehat{\mathbf{g}}_{{\rm{C}},k} \buildrel \Delta \over= \mathbf{E}_{{\rm{C}},k}^{-\frac{1}{2}}\Delta \mathbf{g}_{{\rm{C}},k}$. The sufficient conditions for the communications SINR constraint \eqref{p5-c} are expressed as 
\begin{align}\label{com_sinr_pe_cond}
&\text{Tr}\left(\mathbf{E}_{{\rm{C}},k}^{\frac{1}{2}} \mathbf{\Psi}_{k}\mathbf{E}_{{\rm{C}},k}^{\frac{1}{2}}\right) - \sqrt{2\ln\left(\frac{1}{\rho_{{\rm{C}},k}}\right)}x_{k}^{C}  + \ln\left(\rho_{{\rm{C}},k}\right)y_{k}^{C} + \nonumber\\
&\widetilde{\mathbf{h}}_{{\rm{C}},k}^{H} \mathbf{\Psi}_{k} \widetilde{\mathbf{h}}_{{\rm{C}},k}  -  \Gamma_k {\sigma}^2_{{\rm{C}},k}  \ge  0,	\\
&\sqrt{ \left\| \mathbf{E}_{{\rm{C}},k}^{\frac{1}{2}} \mathbf{\Psi}_{k}\mathbf{E}_{{\rm{C}},k}^{\frac{1}{2}} \right\|_{\text{F}}^{2} + 2\left\| \mathbf{E}_{{\rm{C}},k}^{\frac{1}{2}} \mathbf{\Psi}_{k}\widetilde{\mathbf{h}}_{{\rm{C}},k} \right\|^{2}}  \le  x_{k}^{C}, \nonumber \\
&y_{k}^{C}\mathbf{I}  +  \mathbf{E}_{{\rm{C}},k}^{\frac{1}{2}} \mathbf{\Psi}_{k}\mathbf{E}_{{\rm{C}},k}^{\frac{1}{2}}  \succeq  \mathbf{0}, y_{k}^{C}  \ge  0,\forall k  \in  \mathcal{K}, \nonumber 
\end{align}
where $x_{k}^{C}$ and $y_{k}^{C}$ are the introduced auxiliary variables.

For the probabilistic covertness constraint \eqref{p5-d}, we define $\Delta \widehat{\mathbf{h}}_{{\rm{W}},i} \buildrel \Delta \over= \mathbf{E}_{{\rm{W}},i}^{-\frac{1}{2}}\Delta \mathbf{h}_{{\rm{W}},i}, \forall i \in \mathcal{T}$ such that $\Delta \widehat{\mathbf{h}}_{{\rm{W}},i} \sim \mathcal{CN}\left(\boldsymbol{0},\mathbf{I}_{M_t}\right)$. 
Similarly, the sufficient conditions are presented as 
\begin{align}\label{covertness_constraint_pe}
&\text{Tr}\left(-\mathbf{E}_{{\rm{W}},i}^{\frac{1}{2}}\mathbf{\Xi}\mathbf{E}_{{\rm{W}},i}^{\frac{1}{2}}\right)-\sqrt{2\ln\left(\frac{1}{\rho_{{\rm{W}},i}}\right)}x_{i} +\ln\left(\rho_{{\rm{W}},i}\right)y_{i}+\nonumber\\
&\eta{\sigma}^2_{{\rm{W}},i}-\widetilde{\mathbf{h}}_{{\rm{W}},i}^{H}\mathbf{\Xi}\widetilde{\mathbf{h}}_{{\rm{W}},i} \ge 0,	\\
&\sqrt{\left\|\mathbf{E}_{{\rm{W}},i}^{\frac{1}{2}}\mathbf{\Xi}\mathbf{E}_{{\rm{W}},i}^{\frac{1}{2}}\right\|_{\text{F}}^{2}+2\left\|\mathbf{E}_{{\rm{W}},i}^{\frac{1}{2}}\mathbf{\Xi}\widetilde{\mathbf{h}}_{{\rm{W}},i}\right\|^{2}} \le x_{i}, \nonumber \\
&y_{i}\mathbf{I} - \mathbf{E}_{{\rm{W}},i}^{\frac{1}{2}}\mathbf{\Xi}\mathbf{E}_{{\rm{W}},i}^{\frac{1}{2}}\succeq \mathbf{0}, y_{i} \ge 0,\forall i \in \mathcal{T},\nonumber
\end{align}
where $x_{i}$ and $y_{i}$ are the introduced auxiliary variables.

\begin{table*}[t]\small
\caption{Arithmetic complexities of proposed designs}
\vspace{-2mm}
\centering
\label{complexity}
\vspace{-4mm}
\begin{threeparttable}
\begin{center}
\renewcommand{\arraystretch}{1.1}
\scalebox{0.95}{
\begin{tabular}{|cl|llll|}
\hline
\multicolumn{2}{|c|}{\multirow{2}{*}{\bf{Proposed ADC designs}}}       & \multicolumn{4}{c|}{\bf{Arithmetic complexity}}                                                                                                                                                               \\ \cline{3-6} 
\multicolumn{2}{|c|}{}                             & \multicolumn{2}{c|}{Transmitter side}                                                                                  & \multicolumn{2}{c|}{Receiver side}                                              \\ \hline
\multicolumn{1}{|c|}{\multirow{2}{*}{\bf{Algorithm 2}}} & First check  & \multicolumn{2}{l|}{$\mathcal{O}\left(\left(\left(K_{o}+K_{c}+1\right)M_t^{2}+K_{o}+K_{c}+4T\right)^{3.5}\right)$}   & \multicolumn{2}{l|}{$\mathcal{O}\left(\left(2T M_r^{2}+2T\right)^{3.5}\right)$} \\ \cline{2-6} 
\multicolumn{1}{|c|}{}                             & Second check & \multicolumn{2}{l|}{$\mathcal{O}\left(\left(\left(3K_{o}+3K_{c}+1\right)M_t^{2}+2K_{o}+2K_{c}+4T\right)^{3.5}\right)$} & \multicolumn{2}{l|}{$\mathcal{O}\left(\left(6T M_r^{2}+4T\right)^{3.5}\right)$} \\ \hline
\multicolumn{1}{|c|}{\multirow{2}{*}{\bf{Algorithm 3}}} & First check  & \multicolumn{2}{l|}{$\mathcal{O}\left(\left(\left(K_{o}+K_{c}+1\right)M_t^{2}+2K_{o}+2K_{c}+8T\right)^{3.5}\right)$}  & \multicolumn{2}{l|}{$\mathcal{O}\left(\left(2T M_r^{2}+4T\right)^{3.5}\right)$} \\ \cline{2-6} 
\multicolumn{1}{|c|}{}                             & Second check & \multicolumn{2}{l|}{$\mathcal{O}\left(\left(\left(3K_{o}+3K_{c}+1\right)M_t^{2}+3K_{o}+3K_{c}+8T\right)^{3.5}\right)$} & \multicolumn{2}{l|}{$\mathcal{O}\left(\left(6T M_r^{2}+6T\right)^{3.5}\right)$} \\ \hline
\end{tabular}
}
\end{center}
\end{threeparttable} 
\vspace{-6mm}
\end{table*}

Above all, by omitting the rank-one constraints \eqref{p3-f} and \eqref{p3-g}, \eqref{p5} is transformed into the following problem 
\begin{eqnarray}\label{p6}
\begin{aligned} 
&\mathop {\max }\limits_{\mathbf{W}_{k},  \mathbf{R}, \mathbf{F}_{1,i}, \mathbf{F}_{0,i},x_{k}^{C},y_{k}^{C},x_{1,i}^{R},x_{0,i}^{R},y_{1,i}^{R},y_{0,i}^{R},x_{i},y_{i},t} \;t \\
{\rm{s.t.}}\; 
&\eqref{rad_sinr_pe_cond},\eqref{com_sinr_pe_cond},\eqref{covertness_constraint_pe},\eqref{p3-b},\eqref{p3-c},\eqref{p3-d},\eqref{p3-e}.
\end{aligned}	
\end{eqnarray}
Similar to \eqref{p4}, \eqref{p6} is still non-convex due to the coupled transmit and receive variables as well as the objective value $t$. Besides, \eqref{p6} is also a convex feasibility-checking problem with given transmit (receive) variables and $t$. By straightforwardly formulating the corresponding transmitter and receiver-side feasibility-checking subproblems, the AO-based double-checking framework as in Algorithm \ref{algo 2} can be readily used to solve the outage-constrained transceiver design under probabilistic error model. To avoid redundancy, we do not repeat the detailed steps and directly refer to them as Algorithm 3 which outputs the optimized radar covariance matrix $\mathbf{R}$, and the rank-one transmit and receive variables $\mathbf{W}_{k}$, $\mathbf{F}_{1,i}$, and $\mathbf{F}_{0,i}$. Then, we further utilize ED to obtain high-quality transmit and beamforming vectors.

The convergence of the proposed designs are determined by the AO-based algorithms. We only need to show the convergence of Algorithm \ref{algo 2}. The convergence of Algorithm 3 can be verified in the same manner. For notational convenience, we denote the transmit and receive variables as $\mathbf{W}\buildrel \Delta \over=[\mathbf{W}_{1},\ldots,\mathbf{W}_{k},\mathbf{R}]$ and $\mathbf{F}\buildrel \Delta \over=[\mathbf{F}_{1,1},\ldots,\mathbf{F}_{1,T},\mathbf{F}_{0,1},\ldots,\mathbf{F}_{0,T}]$, respectively. Then, the objective value ${t}$ can be regarded as a function w.r.t. $\mathbf{W}$ and $\mathbf{F}$, i.e., ${t}(\mathbf{W},\mathbf{F})$. In each iteration of Algorithm \ref{algo 2}, the objective value is updated sequentially by the loops w.r.t. $\mathbf{W}^{\left(l\right)}$ and $\mathbf{F}^{\left(l\right)}$, respectively. In particular, for a given $\mathbf{F}^{\left(l-1\right)}$, any $t^{\left(l\right)}$ enumerated by bisection with double-checked feasible $\mathbf{W}^{\left(l\right)}$ is greater than the initial value of $t^{\left(l\right)}$ in this transmitter-side loop. For a given $\mathbf{W}^{\left(l\right)}$, any $t^{\left(l\right)}$ enumerated by bisection with double-checked feasible $\mathbf{F}^{\left(l\right)}$ is greater than the initial value of $t^{\left(l\right)}$ in this receiver-side loop. Hence we have ${t}(\mathbf{W}^{\left(l-1\right)},\mathbf{F}^{\left(l-1\right)}) \le {t}(\mathbf{W}^{\left(l\right)},\mathbf{F}^{\left(l-1\right)}) \le {t}(\mathbf{W}^{\left(l\right)},\mathbf{F}^{\left(l\right)})$. We notice that ${t}(\mathbf{W}^{\left(l\right)},\mathbf{F}^{\left(l\right)})$ is upper-bounded by $\overline{t}$. As a result, the convergence of Algorithm \ref{algo 2} can be guaranteed. Moreover, as $l \to +\infty$, it can be shown that Algorithm \ref{algo 2} converges to a stationary point of \eqref{p4}\cite{razaviyayn2013unified}.

The arithmetic complexities of the proposed ADC designs consist of the complexities of the first and second checks of the transmitter and receiver-side designs, respectively. We present the component-wise complexities for both Algorithms 2 and 3 in Table \ref{complexity}. Let $C_{\text{1,tx}}$, $C_{\text{2,tx}}$, $C_{\text{1,rx}}$, and $C_{\text{2,rx}}$ denote the complexities of the first and second checks of the transmitter and receiver-side designs.
The worst-case overall complexity of each algorithm can be determined by $l_{\text{tx}}(C_{\text{1,tx}}+J C_{\text{2,tx}})+l_{\text{rx}}(C_{1,\text{rx}} + J C_{2,\text{rx}})$, where $l_{\text{tx}}$ and $l_{\text{rx}}$ denote the corresponding transmitter and receiver-side loop numbers, respectively.

\section{Numerical Results}
In this section, we present the numerical results to evaluate the performance of the proposed algorithms for the covert ISAC system. Unless specified otherwise, the basic simulation parameters are set as follows. The ISAC BS is equipped with $M_t = 6$ transmit antennas and $M_r = 6$ receive antennas with half-wavelength interval. The power budget of the BS is set as $P=30$dBm. There are $T=2$ targets located at $80^{\circ}$ and $100^{\circ}$, respectively, and $C=2$ clutters located at $40^{\circ}$ and $150^{\circ}$, respectively. Each of the radar channel coefficients are given by $\alpha_j=0$dB and $\beta_i=0$dB. Besides, the BS serves $K_{o}=2$ overt users and $K_{c}=2$ covert users. Each communications channel is assumed to be Rayleigh fading and $\mathbf{h}_{{\rm{C}},k} \sim \mathcal{CN}\left(\boldsymbol{0}_{M_t},\mathbf{I}_{M_t}\right)$. The noise powers are set as ${\sigma}^2_{{\rm{W}},i}=0$dBm, ${\sigma}^2_{{\rm{C}},k}=0$dBm, and ${\sigma}^2_{{\rm{R}}}=0$dBm. The blocklength is set as $N=1000$ while the covertness constant is set as $\epsilon = 0.1$. The overt and covert communications SINRs are set as $\Gamma_{1}=\Gamma_{2}=2$dB and $\Gamma_{3}=\Gamma_{4}=2$dB, respectively. 
The outage probabilities are set as $\rho_{{\rm{R}},1}\ldots=\rho_{{\rm{R}},1}=\rho_{{\rm{R}}}=0.05$, $\rho_{{\rm{W}},i}=0.05$, $\rho_{{\rm{C}},k}=0.05$.
The CSI uncertainty is controlled by a coefficient given by $\kappa= 0.01$.
To present a fair comparison between the worst-case and outage-constrained designs, we follow the settings in \cite{gui2020tsp}. For the probabilistic error model, we set $\mathbf{E}_{{\rm{C}},k}=\kappa \|\widetilde{ \mathbf{h}}_{{\rm{C}},k}\|^{2}\mathbf{I}_{M_t}/{M_t}$, $\mathbf{E}_{{\rm{W}},i}=\kappa \| \widetilde{\mathbf{h}}_{{\rm{W}},i}\|^{2}\mathbf{I}_{M_t}/{M_t}$, and $\mathbf{E}_{{\rm{R}}}=\kappa \|\widetilde{\mathbf{g}}_{\rm{R}}\|^{2}\mathbf{I}_{(T+C)M_t M_r}/((T+C)M_t M_r)$, respectively.
Accordingly, for the bounded error model, we set $e_{{\rm{C}},k}^{2} =\frac{ \kappa \| \widetilde{\mathbf{h}}_{{\rm{C}},k}\|^{2}}{2}F_{{\chi}_{2M_{t}}^{2}}^{-1}(1-\rho_{{\rm{C}},k})$, $e_{{\rm{W}},i}^{2}=\frac{  \kappa \|\widetilde{\mathbf{h}}_{{\rm{W}},i}\|^{2}}{2}F_{{\chi}_{2M_{t}}^{2}}^{-1}(1-\rho_{{\rm{W}},i})$, and $e_{{\rm{R}}}^{2} =\frac{  \kappa \|\widetilde{\mathbf{g}}_{\rm{R}}\|^{2}}{2}F_{{\chi}_{2(T+C)M_t M_r}^{2}}^{-1}(1-\rho_{{\rm{R}}})$,
respectively. For conciseness, in the following figures, results obtained via Algo. 2 under bounded error model and via Algo. 3 under probabilistic error model are referred as results under BE and PE, respectively.

\begin{figure*}
\centering
\includegraphics[width=6.8 in]{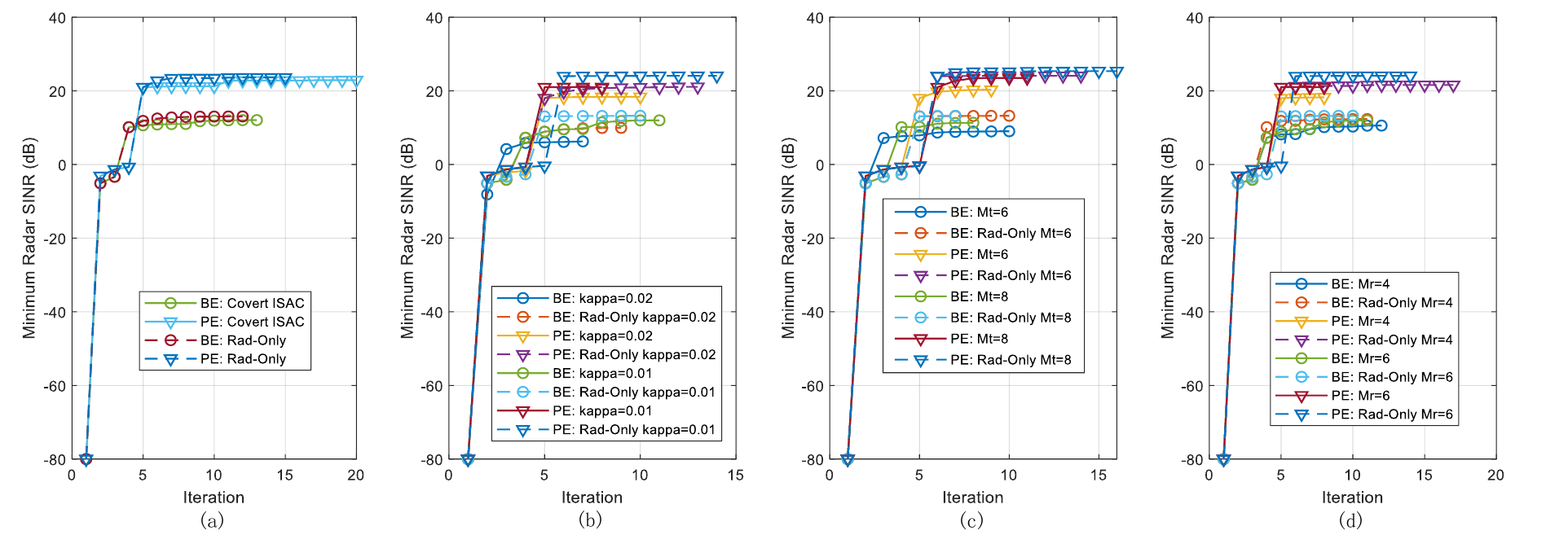}
\vspace{-4mm}
\caption{Illustration of algorithmic convergence. (a) Comparison between the considered covert ISAC system and the radar-only upper bound. (b) Different uncertainty coefficients. (c) Different transmit antenna numbers. (d) Different receive antenna numbers.
}
\label{conv}	
\end{figure*}

\begin{figure*}
\centering
\includegraphics[width=6.8 in]{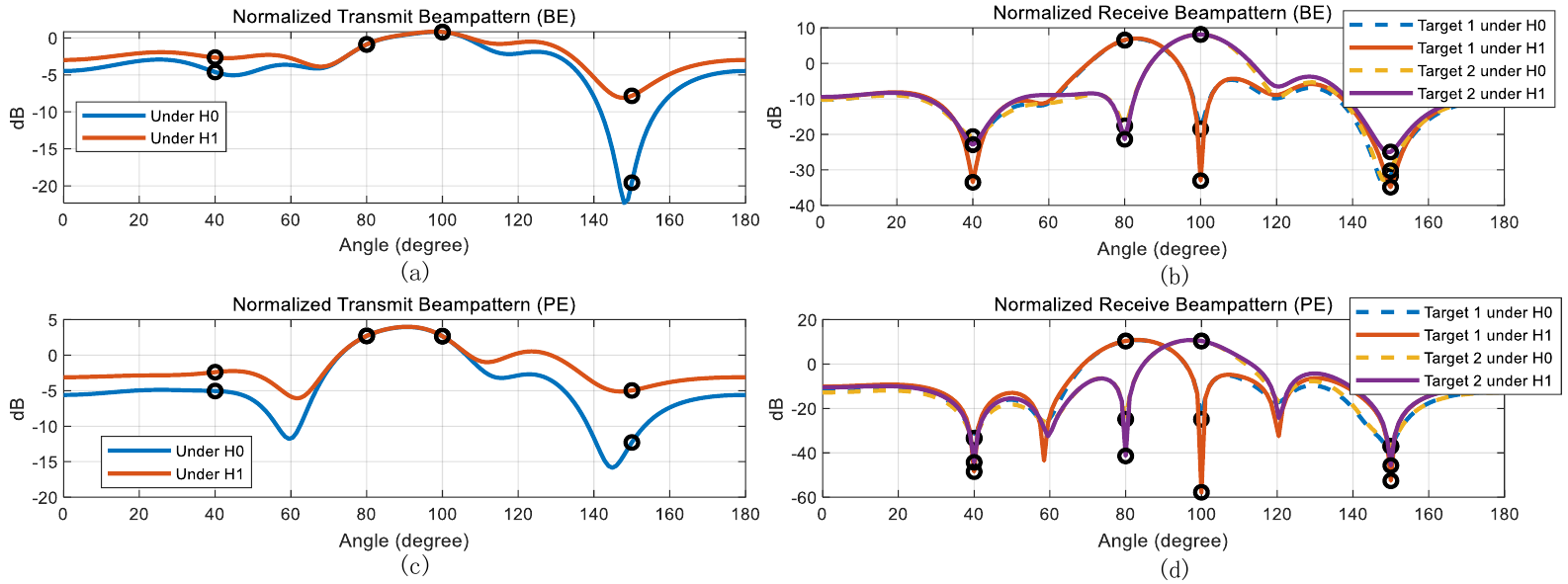}
\vspace{-4mm}
\caption{Beampatterns. (a) Transmit beampattern under BE. (a) Receive beampattern under BE. (c) Transmit beampattern under PE. (d) Receive beampattern under PE. 
}
\vspace{-6mm}
\label{bp}	
\end{figure*}

Figures \ref{conv}(a)-(d) depict the convergence trends of the proposed algorithms compared to their radar-only upper bounds for different uncertainty coefficients and transmit/receive antenna numbers. The iteration number accumulates both the transmitter and receiver-side iteration. We can see that, all of them under different parametric configurations converge very fast. Furthermore, for both results under BE and PE, less than 3 transmitter-receiver cycles of the proposed ADC framework are needed for the algorithms to converge to the stationary points regardless of the different antenna numbers and channel uncertainties. The radar SINR can be boosted significantly within the first two iterations. Meanwhile, higher radar SINR is obtained with less CSI uncertainty and more transmit and receive antennas, i.e., less systematic inaccuracy and more spatial DoFs. As validated in Figs. \ref{conv}(a)-(d) and also in following simulations, owing to more conservative constraints, the worst-case design under BE leads to lower radar SINR than the outage-constrained counterpart under PE.

Figures \ref{bp}(a)-(d) illustrates the transmit and receive beampatterns under both $\mathcal{H}_1$ and $\mathcal{H}_0$. In these subfigures, Figs. \ref{bp}(a) and (b) are the transmit and receive beampatterns under BE, respectively, while Figs. \ref{bp}(c) and (d) are the transmit and receive beampatterns under PE, respectively. As can be seen, in two phases, the transmit beampatterns do not necessarily have good resolutions towards the directions of the targets. However, after the receive beamforming, the receive powers at the directions of the targets are substantially boosted while those of the clutters are deeply nulled. In addition, for the receive beamforming w.r.t. target 1, the direction w.r.t. target 2 is also nulled and vice versa. As a result, the radar SINRs for both targets can be significantly improved after the receive beamforming of the proposed algorithms. The insight here is that it is not necessary to approximate the ideal transmit beampattern, as in \cite{fan2018twc,fan2018tsp,xiang2020tsp,rang2021jstsp,xiang2022jsac}, to obtain desirable radar performance. This is because transmit beampattern is just an transmitter-side intermediate result instead of the overall measurement for radar performance. The transmitter-side-only designs fail to exploit the receiver-side potentials to further enhance the radar performance.

\begin{figure*}
\centering
\includegraphics[width=6.8 in]{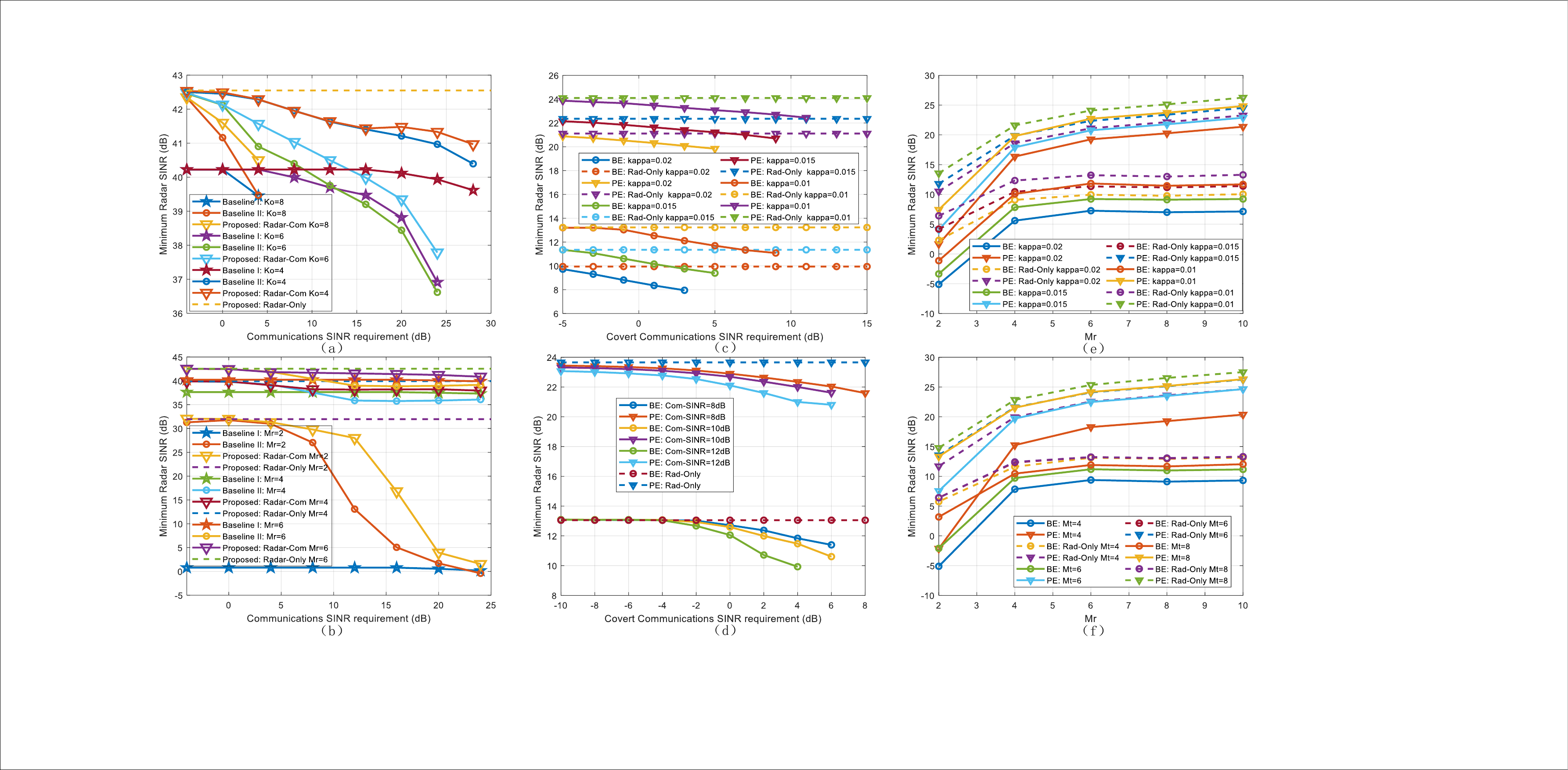}
\vspace{-4mm}
\caption{(a) and (b) are the minimum radar SINR versus communications SINR requirement with perfect CSI for different overt user numbers and receive antenna numbers, respectively. (c) and (d) are the minimum radar SINR versus covert communications SINR for different uncertainty coefficients and overt communications SINRs, respectively. (e) and (f) are the minimum radar SINR versus receive antenna numbers for different uncertainty coefficients and transmit antenna numbers, respectively. 
}
\vspace{-6mm}
\label{total}	
\end{figure*}




To showcase the efficacy and versatility of the ISAC transceiver design framework, we evaluate its performance in the overt communications-only scenario with perfect CSI. In Figs. \ref{total}(a) and (b), we present the results of applying the proposed scheme and compare the minimum radar SINR with two state-of-the-art baselines \cite{xiang2020tsp,yonina2023arxiv}, both designed specifically for scenarios with perfect CSI. The design goals for \cite{xiang2020tsp} (Baseline I) and \cite{yonina2023arxiv} (Baseline II) are summarized as minimizing the mean square error of the MIMO radar beampattern and maximizing the minimum radar SINR with receive beamforming matching the steering vector of the receive antenna array, respectively. Figures \ref{total}(a) and (b) demonstrate that our approach significantly outperforms the two baselines, especially when the communications SINR is relatively high, the number of served users are relatively large, and the receiver antenna numbers are relatively small. The reason for this is that our scheme fully exploits the ISAC transceiver by alternately designing the transmitter and receiver while Baseline I ignored the receiver-side design and Baseline II did not design the radar receiver adaptively.


Figures \ref{total}(c) and (d) plot the minimum radar SINR versus the covert communications SINR for different uncertainty coefficient $\kappa$ and overt communications SINRs, respectively.
In Fig. \ref{total}(c), we can find that for a given covert communications SINR, the radar SINR decreases as $\kappa$ becomes larger. 
The is because as the CSI uncertainty increases, the BS needs to consume more resources to compensate for this inaccuracy, so that the resources used for sensing the actual radar targets gradually decrease. 
In Fig. \ref{total}(d), we can see that the radar SINR decreases as the the covert communications SINR requirement increases for both type of designs. Besides, for a same level of radar SINR, the maximum achievable covert communications SINR decreases with the overt communications SINR. These clearly reveal the tripartite trade-off between the radar and overt/covert communications performances in the covert ISAC system. Moreover, we can observe that when the covert communications SINR is sufficiently small, the minimum radar SINR of the covert ISAC system can approach that of the system which only serves as a MIMO radar. This is due to the transmitted waveform is consist of not only the communications waveform but also the dedicated radar waveform which endows it abundant DoFs to approximate the ideal radar-only covariance, especially when the communications performance requirement is not critical. 

Figures \ref{total}(e) and (f) show the minimum radar SINR versus the receive antenna number $M_r$ with different coefficient $\kappa$ and the transmit antenna number $M_t$, respectively. We can see that as receive antenna number further increases, the radar SINR gain will diminish. This is because the receive antenna number no longer serves as the primary limiting factor for the radar performance.
In practice, we should carefully decide the antenna number to achieve a better trade-off between performance and hardware cost. 
From Fig. \ref{total}(e), we can observe that the uncertainty worsens the radar performance. For a given $M_r$, the radar SINR decreases as $\kappa$ becomes larger, which is in line with the findings in Fig. \ref{total}(c). From Fig. \ref{total}(f), we can see that for a given $M_r$, the radar SINR increases as $M_t$ becomes larger. In addition, the radar performance gap between the different transmit antenna numbers will shrink as $M_r$ becomes larger. This is because the receiver-side DoFs can compensate for the insufficient transmitter-side DoFs, which further highlights the importance of receive beamforming design in covert ISAC systems..


\section{Conclusion}

In this paper, we presented an optimization framework for the robust transceiver design in a covert ISAC system under imperfect CSI. Our approach balanced radar performance, communications requirements, and covertness of the system in the face of challenging non-convex optimization problems involving SICs and coupled beamforming vectors. To overcome these issues, we introduced the S-procedure, Bernstein-type inequality, and ADC robust optimization framework to facilitate feasibility checking and optimization of the transceiver beamforming vectors.
Numerical results revealed the superiority of the proposed ISAC transceiver design framework, compared to the state-of-the-art designs, and demonstrated its robustness. Moreover, the significance of exploiting the receiver-side potentials on enhancing the performance of covert ISAC systems was highlighted.

\bibliographystyle{IEEEtran}
\bibliography{IEEEabrv,mybib}


\end{document}